%% file: main.tex
\documentclass[a4paper,twocolumn,11pt,unpublished]{quantumarticle}
\pdfoutput=1

\usepackage[utf8]{inputenc}
\usepackage[english]{babel}
\usepackage{csquotes}
\usepackage[stable]{footmisc}
\usepackage{hyperref}

\usepackage{amsmath, amsfonts, amssymb}
\usepackage{graphicx}

\usepackage{caption}
\usepackage{subcaption}
\usepackage{adjustbox}
\usepackage{wrapfig}
\usepackage{booktabs}
\usepackage{nicefrac}
\usepackage{microtype}
\usepackage{xcolor}
\usepackage{lipsum}

\title{Optimizing Quantum Photonic Integrated Circuits using Differentiable Tensor Networks}

\author{Mathias Van Regemortel}
\affiliation{Large-Scale Integrated Photonics Lab, Hewlett Packard Labs, HPE Belgium, Diegem, Belgium}
\email{mathias.van-regemortel@hpe.com}
\author{Thomas Van Vaerenbergh}

\renewcommand{\vec}[1]{\mathbf{#1}}

\begin{document}

\maketitle
\begin{abstract}
Recent reports of large photonic nonlinearities in integrated photonic devices, using the strong excitonic light-matter coupling in semiconductors, necessitate a tailored design framework for quantum processing in the limit of low photon occupation. We present a gradient-based optimization method for quantum photonic integrated circuits, which are composed of nonlinear unitary coupling gates and stochastic, nonunitary components for sampling the photonic losses. As core of our method, differentiable tensor-networks are leveraged, which are accurate in the regime of low photonic occupation and modest intermode entanglement. After characterizing the circuit gate architecture with field simulations of GaAs-based samples, we demonstrate the applicability of our method by optimizing quantum photonic circuits for two key use cases: integrated designs for quantum optical state preparation and tailored optimal readout for quantum phase sensing.
\end{abstract}

\input{0-intro/introduction}

\input{1-field-equation/field-equation}

\input{2-tensor-optimization/optimization}

\input{3-stategen/stategen}

\input{4-metrology/metrology}

\input{5-conclusions-outlook/conclusions_outlook}

\section*{Acknowledgements}
We thank Wolfger Peelaers for insightful discussions and a careful reading of the manuscript. We are also grateful to Michiel Wouters for valuable input, particularly on polaritonics, and to Luis Pedro Garcia-Pintos for his contributions on quantum sensing, both of whom also carefully read the manuscript. We further acknowledge our partners in the project “Quantum Optical Networks based on Exciton-polaritons” (Q-ONE) for stimulating discussions during the development of this work. This project has received funding from the European Union’s EIC Horizon Europe Pathfinder Challenges program under grant agreement No. 101115575 (Q-ONE). The views and opinions expressed are those of the authors only and do not necessarily reflect those of the European Union, the European Innovation Council, or the SMEs Executive Agency (EISMEA). Neither the European Union nor the granting authority can be held responsible for them.

\appendix
\input{appendix/appendix}

% \bibliographystyle{unsrt, abbreviate} 
% \bibliography{refs}

\bibliographystyle{quantum}
\bibliography{refs}

\end{document}

%% file: 0-intro/introduction.tex
% \begin{figure}[!h]
%     \centering
%     \includegraphics[width=0.8\linewidth]{0-intro/figs-circuit-optimization.pdf}
%     \caption{The optical circuits used for quantum optimization. A pulsed coherent laser light is coupled into the integrated waveguides and propagates through a bricked pattern of optical nonlinear gates. After the nonlinear quantum processing in the optical chip, the output light is detected for analysis and optimization. The inset illustrates the gate design in more detail; it consists of a linear coupling rate $J$ between two adjacent waveguides, a photonic Kerr-nonlinearity $U$, constant throughout, and a gate propagation time $\Delta t$. The couplings $J$ are iteratively adjusted by simulating the optical quantum circuit and minimizing a figure of merit defined on the output light.}
%     \label{fig: circuit-setup}
% \end{figure}

\begin{figure*}[!ht]
    \centering
    \includegraphics[width=0.6\textwidth]{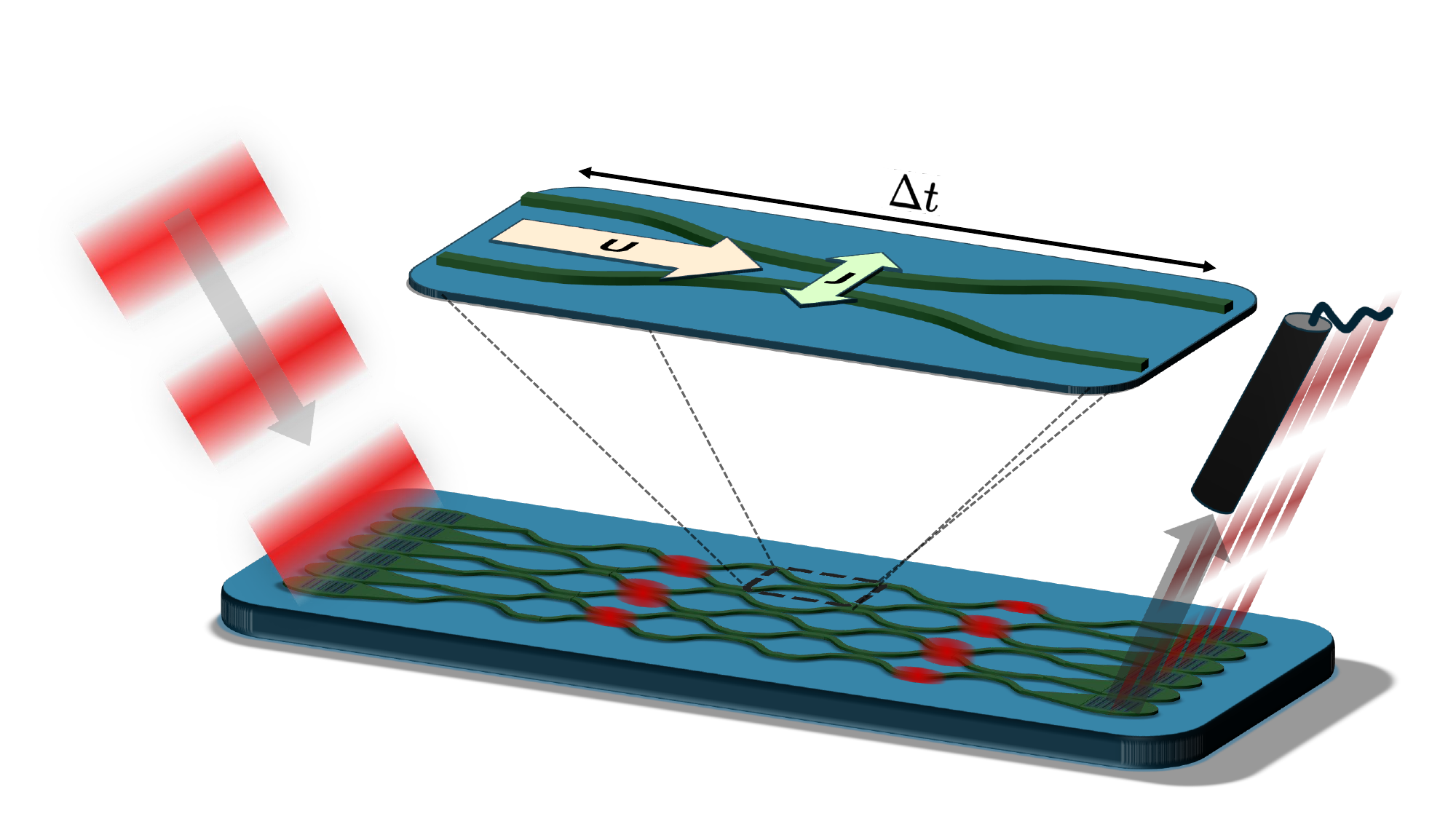}
    \caption{The optical circuits used for quantum optimization. A pulsed coherent laser light is coupled into the integrated waveguides and propagates through a bricked pattern of optical nonlinear gates. After the nonlinear quantum processing in the optical chip, the output light is detected for analysis and optimization. The inset illustrates the gate design in more detail; it consists of a linear coupling rate $J$ between two adjacent waveguides, a photonic Kerr-nonlinearity $U$, constant throughout, and a gate propagation time $\Delta t$. The couplings $J$ are iteratively adjusted by simulating the optical quantum circuit and minimizing a figure of merit defined on the output light.}
    \label{fig:circuit-setup}
\end{figure*}

\section{Introduction}
% start: recent experiments -> high nonlinearity -> quantum statistics/tasks

Recent experiments reported that optical nonlinearities can be impressively enhanced with external electric fields \cite{tsintzos2018electrical, suarez2021enhancement, liran2024electrically}, opening the way to quantum processing using the strong light-matter excitonic coupling in semiconductor samples \cite{keeling2007collective, carusotto2013quantum}. Planar samples, provide, on one side, the flexibility for etching optical gate architectures in 2D topologies and, on the other, the near-infrared wavelength compatibility that aligns with existing fiber-optic infrastructures. This opens up a pathway for integrating semiconductor quantum processing platforms into state-of-the-art channels for communication and sensing.

When harnessed appropriately, the non-classical photonic statistics may outperform the classical counterpart in both computing, as well as, sensing applications, as was recently shown for several experimental test cases, such as Gaussian boson sampling \cite{zhong2020quantum} and photonic quantum sensing and metrology \cite{cimini2024variational, dowran2024parallel,knoll2025experimental}.
 This transition into quantum regimes is achieved through various mechanisms, such as optical parametric oscillation (OPO) \cite{giordmaine1965tunable,paschotta2008optical}, exciton-mediated Kerr-type nonlinearities \cite{gibbs2011exciton,delteil2019towards}, and even full photon blockade \cite{birnbaum2005photon} —- each offering distinct avenues for accessing and controlling strongly non-classical states of light.

 While numerous optimization frameworks have been developed and benchmarked for the design of classical linear photonic integrated circuits (PICs) to perform tasks in the domain of optical computation \cite{Laporte2019-eo,gu2022,jiang2024adeptzzeroshotautomatedcircuit,markovic2020physics, kazanskiy2022optical, mcmahon2023physics,stroev2023analog} and machine learning \cite{musumeci2018overview}, the systematic capture of relevant statistics for quantum processing in quantum PICs (qPICs) remains largely unexplored. The coherent Poissonian photon statistics, at the heart of linear optics, is broken by the presence of optical Kerr-type nonlinearities, thus enabling the generation of light with strong non-classical photon statistics, beyond the amplitude-phase representation of classical coherent laser light. Instead, the description of the photonic quantum state must be explicitly integrated in the qPIC optimization scheme, which can be achieved by, either, a truncated, but exponentially growing bosonic Hilbert space, a suitable variational ansatz, or phase-space sampling methods \cite{gardiner2004quantum}. 

In this work, we use the matrix product state (MPS) representation of the multimode bosonic photon state, where each mode is modeled by a truncated local Fock space, well-suited in the low photon-occupation regime \cite{lauchli2008spreading,lubasch2018tensor,oh2021classical}. Building on this, we present a gradient-based variational optimization method for qPICs, using the automatic differentiation of tensor contractions \cite{liao2019differentiable,zhang2023tensorcircuit, rogerson2024quantum}.
% In this work, we present a gradient-based variational optimization method for qPICs, based on differentiable tensor contractions within a matrix product state (MPS) representation of the multimode bosonic photon state, where each mode is described by a truncated local Fock space \cite{lauchli2008spreading,lubasch2018tensor,oh2021classical}. 
Two key features underpin the suitability of the method for quantum photonic optimization in this context: (i) the modest, area-law entanglement characteristic of these quantum states can be controllably captured using the matrix product state (MPS) formalism \cite{vidal2003efficient}, and (ii) stochastic waveguide losses can be incorporated via Monte Carlo trajectory sampling \cite{dum1992monte,molmer1993monte,carmichael1993quantum} in the MPS formalism \cite{daley2014quantum, van2021entanglement,vovk2022entanglement,van2022monitoring}, while maintaining the differentiability of tensor operations. This is essential for gradient-based optimization. Next, we demonstrate the possibilities of the proposed optimization method through two key applications in the regime of low photon occupation: (i) the design of optimal qPICs for quantum state preparation and (ii) the achieving of optimal readout for quantum phase sensing.

% Crucially, slight (area-law) quantum state entanglement is controllably contained within this method \cite{vidal2004efficient}, as well as the possibility to sample ensembles of stochastic photonic loss events while tracking gradients. The programmability of the chip architectures can be fully leveraged to optimize specific designs for well-defined tasks related to optical quantum processing. While previously gradient-based optimization methods for quantum circuits have been developed and carefully investigated for various quantum-oriented tasks in Ref. \cite{zhang2023tensorcircuit}, we need a solid baseline for (i) photonic (bosonic) quantum states, (ii) optimization using the design flexibility of Kerr nonlinear coupling gates, (iii) the inclusion of photonic noise channels (losses) using stochastic sampling, and (iv) address specific photonic optimization tasks, here focusing on quantum optical state generation and optimal readout for quantum phase sensing. After thorough consideration and analysis, it appeared more beneficial and efficient for us to develop a fully new open-source codebase, explicitly targeting the needs for quantum optical circuits, rather than starting from an existing one -- see Ref. \cite{coderepo} for the published open-source repository.

\paragraph{Contributions}
 The design of the qPIC is illustrated on Fig.~\ref{fig:circuit-setup}; for the optimization the start is to define a figure of merit (FOM) $\mathcal{L}\Big[|\psi_\text{out}\rangle\Big]$ on the qPIC output state $|\psi_\text{out}\rangle$ for executing the optimization of the targeted quantum processing task. Considering injection of coherent laser pulses into the circuit, the nonclassical circuit output state depends explicitly on the circuit gate couplers, $\big|\psi_\text{out}(\{J_{l,d}\})\big\rangle$, with $\{J_{l,d}\}$ the coupling rate of gate $l$ in circuit layer $d$, which couple the waveguides $l$ and $l+1$. The final objective of the optimization is to obtain a set of optimal couplers,
\begin{equation}
    \big\{J_{l,d}^\text{(opt)}\big\}:= \text{argmin}_{\{J_{l,d}\}}\mathcal{L}\Big[\big|\psi_\text{out}(\{J_{l,d}\})\big\rangle\Big].
\end{equation}
For a given waveguide etch depth on the semiconductor chip, the couplings $J_{l,d}$ can be controlled in the fabrication by varying the coupler separation and length \cite{haus2002coupled, xing2017behavior}.

In this work, we contribute to the development of qPIC-based quantum processing in the following ways,
\begin{enumerate}
    \item A framework for solidifying the circuit gate description and parametrization is presented, based on 2D field simulations of GaAs etched samples -- see Sec.~\ref{sec:field} for a summary and Appendices \ref{app:MF2D}-\ref{app:MF1D} for the details of the simulations.
    \item The development of a differentiable optical tensor framework for qPIC simulation and optimization. While similar methods for quantum optimization were presented before in Ref. \cite{zhang2023tensorcircuit}, we developed our own code repository, specifically tailored for (i) photonic (bosonic) quantum simulation and optimization, (ii) optimization within the manifold of Kerr nonlinear coupling gates, (iii) the inclusion of photonic noise channels using stochastic sampling, and (iv) addressing specific photonic optimization tasks. It is publicly available in Ref. \cite{coderepo} -- see Sec.~\ref{sec:optimization}.
    \item Establishing qPIC designs for quantum state generation. Here, we focus on deep-circuit cat-state generation and noisy single-photon sampling, with inclusion of photonic losses --  see Sec.~\ref{sec:stategen}.
    \item Optimizing a qPIC design for quantum phase-sensing readout from coherent input light -- see Sec.~\ref{sec:metrology}.
\end{enumerate}

The optimization framework is, in principle, platform-agnostic and requires only the dimensionless gate coupling $J\Delta t$ and nonlinearity $U\Delta t$, and the coupler loss fraction $\gamma\Delta t$, with $\Delta t$ the photonic gate propagation time, as inputs for optimization. Nevertheless, we use GaAs-based polariton waveguides because they are an appealing experimental baseline platform \cite{walker2013exciton, dietrich2016gaas, shapochkin2018polarization}. Indeed, GaAs quantum wells, in the strong light-matter coupling regime, host electrically tunable Kerr-nonlinearities spanning up to two orders of magnitude \cite{tsintzos2018electrical,suarez2021enhancement,liran2024electrically}. Moreover, the polariton group velocity and the coherence length are compatible with sub-millimeter gate footprints, with $1$ ps low-intensity pulsed-laser excitation in the near-infrared regime -- about $800$ nm wavelength. These features make GaAs a natural first choice for few-photon qPIC optimization, and we highlight that gate nonlinearities up to $U\Delta t \gtrsim 0.2$ are within reach (see Appendices \ref{app:MF2D} and \ref{app:MF1D}). In practice, this constitutes a difference of about 4 orders of magnitude when compared with other materials, such as silicon (Si), which operates at lower-frequency telecom wavelengths, by default $1550$ nm \cite{dinu2003third}.

It should be noted, though, that while GaAs excitonic single-mode Kerr-nonlinearities under strong coupling exceed any other material at near-infrared wavelengths \cite{vladimirova2010polariton}, there are other challenges for the fabrication and testing procedure that require further attention. First of all, the presence of quantum wells within the GaAs-based heterostructure constrains the maximum achievable etching depth, as etching through the active layer degrades the excitonic quality and introduces surface defects. This imposes a trade-off between transverse photonic confinement and in/out-coupling efficiency on one side, and preservation of the excitonic coherence on the other~\cite{riminucci2022nanostructured}. As a consequence, the limited etch depth results in low-index-contrast and weaker transverse photonic waveguide confinement, demanding larger bend radii and coupler lengths, thus increasing the overall circuit footprint, and compounding photonic losses at bends and during in- and out-coupling. In that regard, increased attention has been given recently to heterogeneous platforms based on SiN for the linear processing (broad wavelength range, strong waveguide signal confinement), in combination with active III-V elements, such as GaAs -- see e.g., \cite{haglund2015silicon}. We will refer to the prospects of implementing the simulated setup in the SiN-GaAs platform a few times throughout the text.

Furthermore, the inherent strong exciton coupling (up to $\sim0.3$ exciton fraction) implies that excitonic losses, including nonradiative recombination and disorder scattering \cite{trifonov2015nontrivial}, must be considered as well. The large part of the work concentrates on developing unitary qPIC optimization schemes, but we will include polaritonic losses in first approximation using the Monte-Carlo trajectory approximation (see Sec. \ref{subsec:single-photon}).

%% file: 1-field-equation/field-equation.tex
\section{The gate design and circuit dynamics from field equations}
\label{sec:field}

As an experimental baseline for implementing the design of the qPIC, we focus on GaAs-based planar heterostructures, which have an exciton resonance close to $780\,\text{nm}$ that mediates the photonic nonlinearity in the strong light-matter coupling regime. The qPIC consists of layers of stacked nonlinear coupling gates, as is illustrated on Fig.~\ref{fig:circuit-setup}. 

Each two mode coupler is defined by an effective coupling Hamiltonian of the form,
\begin{equation}
\label{eq:Hcoupler}
  H^{(2)}(J,U) = -\hbar J \Big(a^\dagger_1 a_{2} + a^\dagger_{2} a_{1}\Big) + \frac{\hbar U}{2}\Big(a_1^{\dagger2}a_1^2 + a_{2}^{\dagger2}a^2_{2}\Big),
\end{equation}
where $J$ describes the tunneling rate, per time, and $U$ the intra-waveguide nonlinearity photons are exposed to, also expressed as a rate (inverse time). When the Hamiltonian \eqref{eq:Hcoupler} is applied for some time $\Delta t$ to the photonic field (the time for photons to transfer through the gate) we retrieve unitary gates of the following form,
\begin{equation}
\label{eq:V2}
    V(J, U; \Delta t) = e^{-i\Delta t\, H^{(2)}(J,U)},
\end{equation}
which are fully characterized by a set of two dimensionless parameters, $(J\Delta t,\, U\Delta t)$, which, for convenience, are assumed constant within the the time span $[t,t+\Delta t]$, when the signal passes through. For hybrid SiN-GaAs heterostructures, it might be convenient to split Eq. \eqref{eq:V2}, into sequential operations, 
\begin{equation}
\label{eq:V2_SiN}
  V(J,U,\Delta t)= V(J, 0;\Delta t_1)\, V(0, U;\Delta t_2)\, V(J, 0;\Delta t_1),  
\end{equation}
with $\Delta t = 2\Delta t_1 + \Delta t_2$, to separate the linear (SiN) and nonlinear (III-V, GaAs) processing.

We consider qPIC designs etched on GaAs-based chips using standard fabrication techniques (see, e.g., Ref.~\cite{riminucci2022nanostructured}) to shape an effective transverse confinement potential for the photonic field. In Appendix~\ref{app:MF2D}, we run 2D field simulations of a nonlinear coupler etched on a planar GaAs sample, excited with a pulsed laser, using realistic experimental parameters. The goal is to identify the parameters $(J,U)$, from Eq.~\eqref{eq:Hcoupler}, that characterize the circuit gates. Next, starting from the 2D field simulations, an effective model of 1D coupled ordinary differential equations (ODEs) can be derived to describe a coupled set of waveguides with coupling parameter $J$. These are subsequently used for gauging the circuit gate parameters $(U,\Delta t)$, considering pulsed laser excitation, close to the exciton resonance. The photons enclosed in one laser pulse in waveguide $l$ can be considered a single bosonic mode $a_l$, in case the spatial extent of the pulse is well-contained, below the polaritonic coherence length. This we verify in Appendix~\ref{app:MF1D} by comparing the photon pulse field dynamics in the waveguide with an effective reduced circuit gate model for the bosonic operator $a_l$.

We give a short overview of the material and pulse parameters that influence the gates.
\begin{itemize}
    \item \textbf{The exciton interaction constant:} The planar GaAs quantum well induces a photonic nonlinearity $g_{2D}$ due to the strong coupling to the interacting excitons \cite{vladimirova2010polariton}. The strength of this interaction sets, for a large part, the effective nonlinearity experienced by the bosonic modes in the circuit representation. Reported values range from $g_{2D} = 6-10 \mu eV\mu m^2$ in unperturbed samples \cite{sun2017direct, suarez2021enhancement}, while applying an electric field gives nonlinearities up to $g_{2D} = 600 \mu eV\mu m^2$, as reported in Ref.  \cite{liran2024electrically}.
    \item \textbf{Polariton group velocity}: the group velocity of the polaritons is determined by their coupling to the exciton field, set by the wavelength and in-plane wavevector of the light. Values of $v_g=26\,\mu\text{m}\,\text{ps}^{-1}$ have been reported in literature \cite{walker2013exciton}. This will, on the one hand, impact $\Delta t$ and, on the other, the spatial extent of the pulse in the sample. More compressed pulses generate higher nonlinearity, as can be seen from,
    \begin{equation}
    \label{eq:U_w}
        U = \int_\mathcal{S} d^2r\, n(\vec{r})\,\big[g_{2D} n(\vec{r})\big],
    \end{equation}
    with $n(\vec{r})$ the normalized photon field density of the pulse in the semiconductor plane $\mathcal{S}$, confined in the transverse electric zero eigenmode of the waveguide ($\text{TE}_0$-mode) for maximal coupling \cite{shapochkin2018polarization}, and $g_{2D} n(\vec{r})$ the nonlinear energy shift at point $\vec{r}$. Thus, lowering $v_g$ amounts to an enhanced effective gate nonlinearity due to an increase in exposure time. This can be achieved at higher exciton fraction or with methods of slow light~\cite{torrijos2021slow}.
    \item \textbf{Excitonic fraction of polariton field:} By varying the light frequency and in-plane wavevector, the exciton fraction of the polariton mode, $a_\text{LP}(\vec{k}) = u_k\,a_\text{ph}(\vec{k}) + v_k\,a_\text{ex}(\vec{k})$, with $u_k^2 + v_k^2=1$, can be scanned with the laser detuning from the exciton resonance, given by $v_k^2$. Higher exciton fraction gives slower propagation (the exciton is largely immobile as compared to the photons) and higher nonlinearity.
    \item \textbf{Etched waveguide separation:} The photonic tunneling rate from one waveguide to the other is, to first order, determined by the overlap between the two transverse waveguide eigenmodes \cite{haus2002coupled}. We assume a conservative separation between the waveguides of the order of $\sim500\,\text{nm}$~\cite{riminucci2022nanostructured}.
\end{itemize}

After analysis in Appendices~\ref{app:MF2D} and \ref{app:MF1D}, we find that a reasonable parameter range for GaAs gates, with a gate length of the order of $100\,\mu \text{m}$, are $0\leq J\Delta t \leq \pi$, maximally a full tunneling rotation between the waveguides, and $0 \leq U \Delta t \leq 0.25$. This is what we will use as a baseline for developing the circuit tensor-based gradient-descent optimization scheme.

%% file: 2-tensor-optimization/optimization.tex
\section{Tensor network circuit optimization}
\label{sec:optimization}

% Having a solid baseline for the architecture of the nonlinear optical gates with modulable coupling rates, we continue with presenting the differentiable tensor network approach for dissipative quantum circuits, valid in the regime of low photon occupation per mode (that is, per waveguide pulse) and limited inter-mode entanglement. This will provide the core of the gradient-based tensor optimization of the qPIC design for quantum tasks. 

% In Sec.~\ref{subsec:circuit}, we first outline the tensor network contraction of the optical circuit using the MPS tensor representation of multimode photonic quantum states. Next, in Sec.~\ref{subsec:MC}, we extend the default MPS representation with an extra batch index, which labels the ensemble of quantum states for stochastically sampling MC trajectories of photonic dissipation with non-unitary gates. Finally, in Sec.~\ref{subsec:optimization-scheme}, we give more details about the gradient-based optimization scheme.

Having established the architecture of nonlinear optical gates with coupling rates as optimization variables that can be determined during the chip-design phase, we now introduce the differentiable tensor network framework for dissipative quantum circuits that allows to find the optimal values of these coupling rates. This approach is valid in the low-photon regime, due to the explicit single-mode bosonic Fock-space truncation, with limited intermode entanglement.

In Sec.~\ref{subsec:circuit}, we describe the tensor network contraction of the optical circuit using the MPS representation of multimode photonic states. In Sec.~\ref{subsec:MC}, we extend this representation with a batch index to enable stochastic sampling of dissipation via non-unitary gates. Finally, Sec.~\ref{subsec:optimization-scheme} details the gradient-based optimization scheme using the tensor network representation of the qPIC.

\subsection{The tensor-network photonic circuit contraction}
\label{subsec:circuit}

\begin{figure}[!ht]
    \centering
    \includegraphics[width=\linewidth]{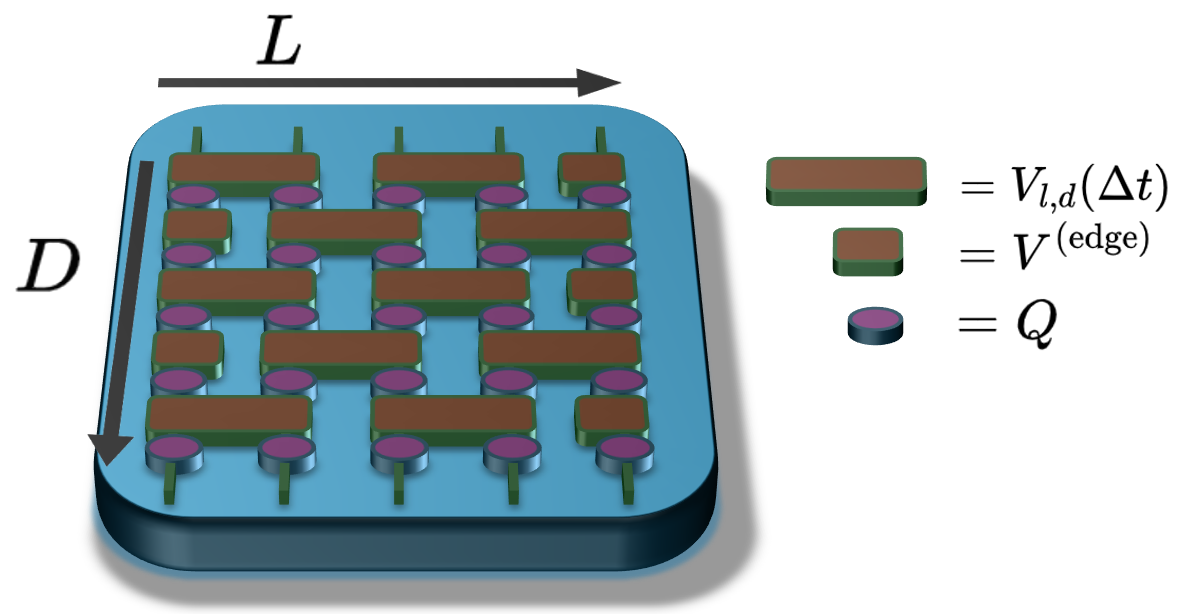}
    \caption{The qPIC setup. The circuit contains $L$ waveguides with unitary coupling gates $V_{l,d}(\Delta t)$ \ref{eq:V2_circ}, stacked in a bricked pattern of layer depth $D$, with unitary edge gates $V^\text{(edge)}$ (brown rectangles). Each layer of unitary gates, is followed by non-unitary stochastic sampling gates $Q$ (Eq.~\eqref{eq:diss-gate}) for the losses (purple dots). }
    \label{fig:circuit}
\end{figure}

The full circuit consists of $L$ parallel waveguides, on which two-mode nonlinear coupling gates are organized in a bricked pattern of depth $D$ -- see Fig.~\ref{fig:circuit}. Each two-mode unitary nonlinear coupling gate $l$ in layer $d$ is defined as,
\begin{equation}
\label{eq:V2_circ}
    V_{l,d} = e^{-i\Delta t_d\, H^{(2)}(J_{l,d},U)},
\end{equation}
with $H^{(2)}(J,U)$ defined in Eq.~\eqref{eq:Hcoupler}, $J_{l,d}$ the gate coupling rate and $U$ the gate nonlinearity, constant across the circuit. The gate time is given as $\Delta t_d = \Delta z_d / v_g$, with $v_g$ the polariton group velocity and $\Delta z_d$ the spatial length of layer $d$.

At the edges of the circuit, where no two-mode gate is overlapping with the edge waveguide, a single-mode unitary gate $V^\text{(edge)}_d = e^{-i\Delta t_d H^\text{(edge)}}$ is applied, with $H^{\text{(edge)}}= \big(U/2\big) a^{\dagger 2} a^2$, thus ensuring a constant intra-waveguide nonlinearity across the circuit.

Unless mentioned otherwise, each waveguide $l$ in the circuit receives as input uncoupled coherent laser light, defined by the complex parameter $\alpha_l$, which exhibits Poissonian photon number statistics, uncorrelated between different modes~$l$,
\begin{equation}
\label{eq:init}
    \big|\psi\big\rangle_\text{in} = \bigotimes_{l=0}^{N}\big|\alpha_l \big\rangle,
\end{equation}
with,
\begin{equation}    
\big|\alpha\big\rangle = e^{\alpha a^\dagger - \alpha^\ast a}\big|0\big\rangle= e^{-\frac{|\alpha|^2}{2}} \sum_i \frac{\alpha^n}{\sqrt{n!}}\big|n\big\rangle.
\end{equation}
Whenever $U>0$, the nonlinearity in the qPIC disrupts the Poissonian photon statistics, giving rise to nonclassical features in the photon number statistics of the output signal.

\begin{figure}[!ht]
    \centering
    \includegraphics[width=\linewidth]{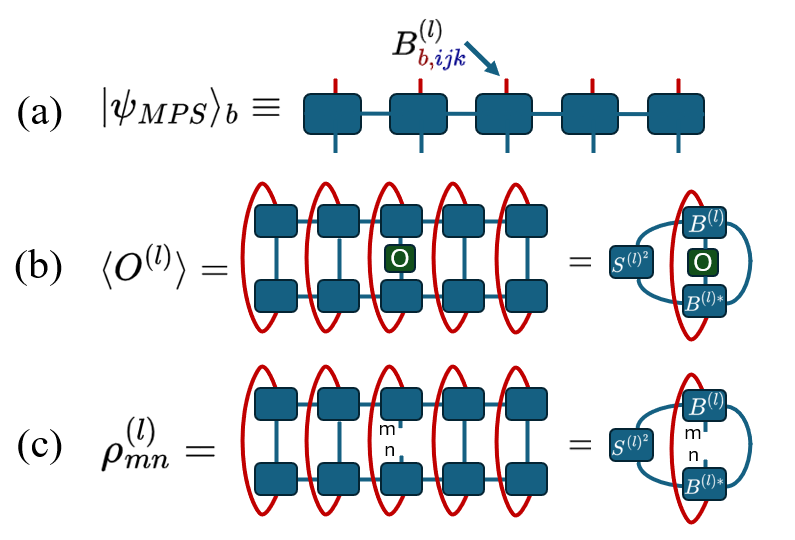}
    \caption{The MPS representation of the MC ensemble of quantum states. (a) Each site tensor is augmented with a fourth index $b$ (red), which labels the batch in the Monte Carlo trajectory ensemble. (b) The tensor contraction for evaluating a local expectation value of a Hermitian operator $O$ in mode (waveguide) $l$; the MPS is contracted with its conjugate, with the $O$ sandwiched in between in mode $l$. (c) The contraction for obtaining the reduced density matrix of mode $l$, $\rho^{(l)}$. (b)-(c) Thanks to the MPS being in right-canonical form, the circuit contraction is local, using the  singular values stored in $S^{(l)}$.}
    \label{fig:MPS}
\end{figure}

\begin{figure*}[!ht]
    \centering  \includegraphics[width=0.8\linewidth]{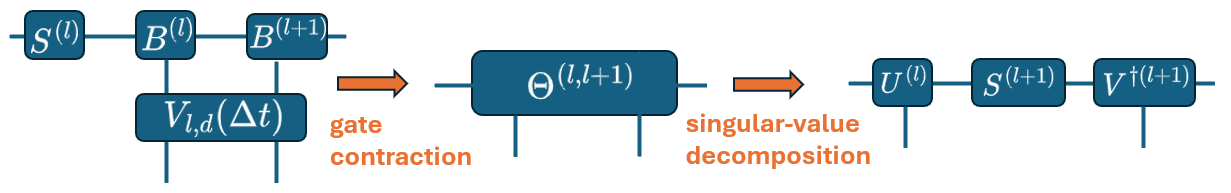}
    \caption{Applying a two-mode unitary gate in layer $d$, $V_{l,d}$, to the MPS state $|\psi_{d-1}\rangle$, exiting layer $d-1$, on mode $l$. First, the unitary gate tensor is contracted with the MPS tensors at site $l$ and $l+1$. Next, the resulting rank-4 tensor $\theta^{(l,l+1)}$ is decomposed into two new rank-3 tensors using SVD. Both the tensor contraction and the SVD decomposition are differentiable tensor operations, of which the gradients can be tracked for evaluating gradients toward circuit parameters after full circuit contraction.}
    \label{fig:unitary-contract}
\end{figure*}
For bosonic states, a truncation $\mathcal{N}_\text{max}$ in Fock number space is needed to represent the photon field in one single mode; this is the physical tensor dimension. The full Hilbert space required for simulating $L$ photonic modes is then $\mathcal{D}=\mathcal{N}_\text{max}^L$, which, as expected, grows exponentially in the number of modes and quickly surpasses computational limits. Another approach is provided by 1D tensor networks (tensor trains); the matrix product state (MPS) representation \cite{cirac2021matrix},
\begin{equation}
\label{eq:MPS}
    |\psi_\text{MPS}\rangle = \sum_{\{n\}} \text{Tr}\big[B_{n_1}^{(1)} B_{n_2}^{(2)}\dots B_{n_L}^{(L)}]\; |n_1n_2\dots n_L\rangle,
\end{equation}
with $|n_1,n_2,\dots, n_L\rangle$ a multimode photon number Fock state. To avoid the cumbersome procedure of writing out all tensor indices, this is often represented as a tensor diagram, as illustrated in Fig.~\ref{fig:MPS}(a).

The quantum state in MPS form \eqref{eq:MPS} is represented by a train of rank-3 tensors $B^{(l)}_{i,j,k}$. Here, the indices $i$ and $k$ the \emph{bond} indices whose dimensions are limited by a chosen cutoff parameter $\chi_\text{max}$, while $j$ labels the truncated Fock space of dimension $\mathcal{N}_\text{max}$ used to represent the photonic mode. The bond indices $i,k$ encode the Hilbert-Schmidt decomposition across a bipartite cut in the tensor chain, and their dimensions, constrained by $\chi_\text{max}$, determine the amount of entanglement that can be faithfully captured within the MPS representation \cite{vidal2003efficient}. For convenience, the MPS is kept in \emph{right-canonical} form, which means that, when indices $(j,k)$ are grouped together, the reshaped matrices $B^{(l)}_{i,(jk)}$ are right-unitary. 

In Fig.~\ref{fig:unitary-contract}, the differentiable method for contracting a circuit gate $V_{l,d}$ with on a photonic MPS state is shown. Contracting a two-mode nonlinear circuit gate $V^{(l,d)}$, represented as a rank-4 tensor, with waveguides $B^{(l)}$ and $B^{(l+1)}$, results in a rank-4 tensor $\Theta^{(l,l+1)}$. Using singular-value decomposition (SVD), the $\Theta^{(l,l+1)}$ tensor is decomposed into two rank-3 tensors, of which the latter is right-unitary. An ordered set of left-singular values $s_{i}^{(l)}$, representing the Schmidt coefficients of a bipartite decomposition across mode $l$, are stored in memory for later use. Iterating from right to left through the chain, contracting the unitary gates $V^{(l,d)}$, results in an updated right-canonical MPS $|\psi_d\rangle$ after application of circuit layer $d$. See Ref. \cite{vidal2003efficient}, in which the circuit contraction method was presented, or Refs. \cite{perez2006matrix, cirac2021matrix} for a full review. The open-source code provided by TeNPy \cite{tenpy2024} served as a reference for implementing the two-mode gate contraction efficiently in PyTorch.

An important note: Although the SVD steps in the circuit contraction are fully differentiable, implementing their gradients for complex-valued matrices (especially when they are close to degeneracy) can be prone to instabilities during the process of gradient backtracking. For the purposes of qPIC optimization, the PyTorch SVD module was overwritten, using the analysis from Ref. \cite{wan2019automatic}, with the necessary amendment of (i) adding explicit SV truncation in the gradient tracking and (ii) regularizing nearly degenerate SVs (the Hilbert-Schmidt coefficients of a bipartite decomposition) -- see source files in \cite{coderepo}.

\subsection{Circuit dissipative losses using Monte Carlo trajectory sampling}
\label{subsec:MC}
The dominant component of circuit dissipation are photonic and excitonic losses, which are caused by material defects and nonradiative exciton recombination  \cite{trifonov2015nontrivial}. We rely on the Monte Carlo quantum trajectory method for stochastically sampling the impact of bosonic loss events on the photonic quantum state. The method was originally presented, independently, in Refs.  \cite{dum1992monte,molmer1993monte,carmichael1993quantum} -- see Ref. \cite{daley2014quantum} for a comprehensive review, which also addresses the implementation with MPS.

In short, dissipative noisy quantum dynamics is generally described by the Lindblad master equation, which acts on the level of a quantum density matrix $\rho$ \cite{gardiner2004quantum},
\begin{equation}
\label{eq:ME}
    \partial_t \rho = \frac{i}{\hbar}\big[\rho, H(t)\big] + \sum_i c_i \rho c_i^\dagger - \frac{1}{2}\Big(c_i^\dagger c_i \rho + \rho c_i^\dagger c_i\Big).
\end{equation}
The first term on the r.h.s. contains the unitary evolution under the time-dependent Hamiltonian $H(t)$ of the statistical ensemble of quantum states, as represented by the density matrix $\rho$. The second term describes the Markovian dissipation generated by the dissipators $c_i$. For photonic losses, these are represented as $c_l\equiv\sqrt{\gamma}\, a_l$, that is, uncorrelated bosonic loss (annihilation) events in waveguide $l$, at a uniform rate $\gamma$. The dissipative part of Eq. \ref{eq:ME} consists of two terms; $\frac{1}{2}\sum_l\big(c_l^\dagger c_l \rho + \rho c_l^\dagger c_l) = \frac{1}{2}\sum_l\big\{c_l^\dagger c_l, \rho\big\}$, with $\{\cdot\}$ being the anticommutator, can, equivalently, be absorbed in the first term by defining a non-Hermitian Hamiltonian $\tilde{H} = H - \frac{i}{2}\sum_i c^\dagger_i c_i$. The quantum noise is described by the dissipative \emph{jump} term $\sum_i c_i\rho c^\dagger_i$, which can be, either, integrated deterministically in the master equation \eqref{eq:ME}, or, alternatively, be sampled stochastically using an ensemble of pure-state Monte Carlo trajectories $|\psi^{(b)}(t)\rangle$. The latter option is the method we use in this work.

The Monte Carlo procedure is as follows; $N_s$ trajectory states $\big|\psi^{(b)}(t=0)\big\rangle$, labeled with batch index $b$, are initiated -- each initial batch state is identical if the system starts from a pure state, as holds for uncoupled coherent laser light described in Eq.~\eqref{eq:init}. Next, each differential time step $\Delta t$, (i) the state $|\psi^{(b)}\rangle$ is unitarily evolved with $H(t)$ and (ii) the probability of a dissipative ``click'' is evaluated for a photon loss event in waveguide~$l$,
\begin{equation}
\label{eq:Pmcclick}
    \Delta p^{(b)}_l(t) = \gamma\Delta t\,n_l^{(b)}(t),
\end{equation}
with $n_l^{(b)}(t) = \langle\psi^{(b)}|a^\dagger_l a_l|\psi^{(b)}\rangle$. The probability $\Delta p^{(b)}_l(t)$ is, as is expected for Poissonian loss statistics, to first order proportional to (i) the rate and time step, $\gamma \Delta t$, and (ii) the photonic occupation at time $t$ of ensemble batch state $b$ in waveguide $l$, $n^{(b)}_l(t)$. 

After drawing a uniform random number $0\leq r^{(b)}_l \leq 1$, there are two possible outcomes,
\begin{itemize}
    \item $r^{(b)} \leq \Delta p^{(b)}_l(t)$: Apply the jump, $\big|\psi^{(b)}(t)\big\rangle \leftarrow a_l\big|\psi^{(b)}(t)\big\rangle$. Physically, this represents the detection of an emitted photon by an observer in time interval $[t,t+\Delta t]$ and, henceforth, a corresponding photon subtraction inside the waveguide $l$.
    \item $r^{(b)} > \Delta p^{(b)}_l(t)$: Perform a non-unitary time step, $\big|\psi^{(b)}(t)\big\rangle \leftarrow e^{-\frac{\gamma\Delta t}{2} a^\dagger_l a_l }\big|\psi^{(b)}(t)\big\rangle$. Physically, this step encodes the information gathered by an external observer by \emph{not} detecting an emitted photon in time interval $[t,t+\Delta t]$. This makes the observer assume that $|\psi^{(b)}\rangle$ is in a state of lower photon occupation in waveguide $l$, as represented by the exponential suppression of higher photon amplitudes.
\end{itemize}
Both outcomes are, in general, non-unitary, implying that the state $|\psi^{(b)}\rangle$ must be renormalized after application, before the next time step can be evaluated.

From the ensemble of trajectory states, the generally non-pure density matrix of the system is reconstructed at each time as,
\begin{equation}
\label{eq:rhoMC}
    \rho(t) \approx \frac{1}{\mathcal{N}_s} \sum_{b=1}^{\mathcal{N}_s} \big|\psi^{(b)}(t)\big\rangle\big\langle\psi^{(b)}(t)\big|
\end{equation}
While the full density matrix is rarely explicitly evaluated, it is important that operator expectation values are, likewise, evaluated from the trajectory ensemble as,
\begin{equation}
\label{eq:OMC}
    \big\langle O(t) \big\rangle \approx \frac{1}{\mathcal{N}_s} \sum_{b=1}^{\mathcal{N}_s} \big\langle \psi_b(t)\big| O \big|\psi_b(t)\big\rangle.
\end{equation}
For local few-body observables, the convergence is generally much faster than the full density matrix \cite{daley2014quantum}.

Within the MPS representation of the quantum state \eqref{eq:MPS}, the default rank-3 tensors, are expanded by adding the ensemble batch index $b$ for representing the ensemble of Monte Carlo trajectory states -- see Fig.~\ref{fig:MPS}(a). A local operator expectation value in waveguide $l$ is evaluated by the contraction of the MPS state with its conjugate, with the local operator $O$ contracted in between, as shown in Fig.~\ref{fig:MPS}(b). The single-mode reduced density matrix of mode $l$ can be obtained in the same way, but now the physical indices of mode $l$ are left open, as shown in \ref{fig:MPS}(c). In all panels, the red indices are the added batch indices of the stochastic ensemble. Contracting the batch indices of a state with the batch indices of its conjugate, provides the ensemble average in the Monte-Carlo trajectory quantum simulation of dissipative dynamics from Eqs.~\eqref{eq:rhoMC}-\eqref{eq:OMC} upon normalization. On the right side of Fig. \ref{fig:MPS}(b)-(c), we illustrate that, since the MPS is kept in right-canonical form, contracting sites right from mode $l$ results in the identity matrix, while left from $l$ requires the contraction of the singular values, which are stored in memory after the two-mode gate contractions. Hence, only a local contraction must be performed to obtain local quantities $\langle O^{(l)}\rangle$ or reduced density matrices $\rho^{(l)}$.

For simulating the stochastic sampling of the Monte-Carlo trajectories using the tensor network representation of the qPIC, a single-mode non-unitary gate is introduced,
\begin{equation}
\label{eq:diss-gate}
    B^{(b)}_{l,kim}\leftarrow \sum_j Q_{l,ij}^{(b)} B^{(b)}_{l,kjm},
    \end{equation}
with,
    \begin{equation}
    Q^{(b)}_l = a \;\text{if}\; r^{(b)}_l \leq \gamma\Delta t \, n^{(b)}_l \;\text{else}\; e^{-\frac{\gamma \Delta t}{2}\, a^\dagger a},
\end{equation}
with $a$ ($a^\dagger$) the bosonic annihilation (creation) operator and $r_b$ an array of uniformly distributed random number between $0$ and $1$, sampled independently for any trajectory state $b$, conform the method presented in Ref.~\cite{molmer1993monte}. Since the stochastic gate $Q^{(b)}_{l,ij}$ is non-unitary, it breaks the right-canonical MPS form, which must be restored after the contraction of Eq.~\eqref{eq:diss-gate}.

When including the dissipation process $Q^{(b)}_{l}$ in Eq.~\eqref{eq:diss-gate}, the computational workload significantly increases because of two reasons, 
\begin{enumerate}
\item A Monte Carlo ensemble needs to be evaluated and sampled, which causes a linear increase in the workload. This can be sampled \emph{embarrassingly parallel} on separate cores, as was previously illustrated in, e.g., Refs.~\cite{van2021entanglement,van2022monitoring}. For this work, however, we prioritize maintaining the tensor structure for efficient gradient tracking. We do this by expanding the default MPS representation with a batch index, as illustrated in Fig.~\ref{fig:MPS}, and use GPU tensor processing provided by the Python library PyTorch~\cite{paszke2019pytorch}. 
\item The non-unitary gates $Q^{(b)}_l$ require the reformatting of the MPS in right-canonical form after application. This involves a sequence of QR and SVD decompositions (see procedure in Ref.~\cite{vidal2003efficient} or the source code from Ref. \cite{tenpy2024}), which, not surprisingly, significantly increases the workload for (i) the circuit evaluation and (ii) the computational steps needed for backtracking the gradients towards the qPIC variables $J_{l,d}$.
\end{enumerate}

\subsection{Gradient-based optimization scheme for qPICs}
\label{subsec:optimization-scheme}

For the optimization, a real-valued differentiable figure of merit (FOM) is defined on the circuit output quantum state $\big|\psi_\text{out}\big\rangle$,
\begin{equation}
    \mathcal{L}(\{J_{l,d}\}):=\mathcal{L}\Big[\big|\psi_\text{out}\big(\{J_{l,d}\}\big)\big\rangle\Big].
\end{equation}
The FOM depends on the set of gate coupling variables $\{J_{l,d}\}$, of which the gradients towards each coupling parameter $\partial \mathcal{L}/\partial J_{l,d}$ are tracked using the autograd provided by PyTorch. More specifically, we use the Adam optimizer \cite{kingma2014adam} to make gradient-based updates of the variables $J_{l,d}^{(i)}$, with $i$ the index of the iteration, starting from some initial condition $J^{(\text{init})}_{l,d}:=J^{(i=0)}_{l,d}$. After iterative circuit contraction and FOM evaluation, $\mathcal{L}^{(i)}$ converges and optimal coupling rates are found; $\{J^{(\text{opt})}_{l,d}\} := \text{argmin}_{\{J_{l,d}\}} \big[\mathcal{L}(\{J_{l,d}\})\big]$. A maximal number of iterations $\mathcal{N}_\text{iter}$ can be set to control the total runtime. 

We note that the Markovian Monte-Carlo trajectory click probabilities $\Delta p^{(b)}_l(t)$ themselves (Eq.~\eqref{eq:Pmcclick}) are not explicitly included in the gradient backtracking algorithm, only the outcomes in terms of the gate tensor contractions $Q_{b,ij}$ of the different batch states (see Eq.~\eqref{eq:diss-gate}). In that regard, we consider variations of $\Delta p^{(b)}_l(t)$ induced by updates of the gate couplings $J_{l,d}^{(i)}$ across subsequent iterations $i$ negligible for the backtracking.

In what follows, we formulate the FOMs $\mathcal{L}$ for two use-cases that have a direct relevance for practical implementation with integrated photonics for near-term quantum processing, quantum state generation (Sec.~\ref{sec:stategen}) and optimal readout of quantum phase sensing (Sec.~\ref{sec:metrology}). In the first case, the goal is to approach a target single-mode quantum state as close as possible in the circuit output mode. In the second, we maximize the qPIC sensitivity towards a small phase shift in one of the incoming coherent light sources.

%% file: 3-stategen/stategen.tex
\section{Quantum state generation}
\label{sec:stategen}

Exciton-polaritons in GaAs structures have emerged as a versatile platform for exploring quantum state generation and manipulation. Their unique light-matter duality enables coherent control and nonlinear interactions, which are crucial for developing quantum technologies. Recent works have demonstrated the potential of exciton-polariton systems in implementing quantum neuromorphic platforms that exploit reservoir computing principles for efficient quantum state preparation, reconstruction, and processing \cite{ghosh2019quantum, ghosh2020reconstructing, ghosh2021quantum}.

The objective for the task of quantum optical state preparation is optimizing the circuit gate architecture for generating a target outcome state, $\rho^\mathrm{(tar)}$, in one circuit mode $l$ (single-mode case) or, more generally, a set of output modes $l_0\dots l_k$ (multi-mode case). Here, we concentrate on the single-mode case, but generalizing the procedure to multi-mode state generation is straightforward.

The starting point for the circuit optimization is the photonic input of uncoupled coherent laser light, as defined in Eq.~\eqref{eq:init}. We set the amplitude uniform across all waveguides, i.e., $\forall l:\;\alpha_l\equiv\alpha$.

\subsection{The figure of merit for state preparation}
\label{subsec:cat}
After contracting the circuit tensor gates, the output MPS is obtained. For the single-mode case, simple contraction of the right canonical MPS gives the single-mode density matrix $\rho^{(l)}_{mn}$ of output mode $l$ -- see Fig.~\ref{fig:MPS}(c). We define the FOM as the weighted trace distance between $\rho^{(l)}$ and the target state $\rho^\mathrm{(tar)}$, defined as follows,
\begin{eqnarray}
\label{eq:Lrho}
    \nonumber\mathcal{L}_\rho &=& T_\mathcal{M}\Big[\rho^{(l)}, \rho^\mathrm{(tar)}\Big]\\
    \nonumber& =& \frac{1}{2}\text{tr}\bigg[\sqrt{\delta \rho^{(l)\dagger}_\mathcal{M} \delta \rho^{(l)}_\mathcal{M}}\bigg]\\ 
    &=& \frac{1}{2} \sum_{r=1}^{\mathcal{N}_\text{max}}|\lambda_r|,
\end{eqnarray}
with $\delta \rho^{(l)}_\mathcal{M} = \mathcal{M}\odot \big(\rho^{(l)} - \rho^\mathrm{(tar)}\big)$ and $\lambda_r$ its eigenvalues. $\mathcal{M}$ is a normalized ``mask'', of which the Hadamard (dot) product gives weights to individual matrix elements of $\rho^{(l)}$ for the optimization scheme. For example, diagonal elements, representing classical probabilities, could be given higher relative weight, or lower weight can be assigned to the vacuum $\rho_{00}^{(l)}$; exceedingly occupying the vacuum can be tolerated, as this limits the generation of spurious nonzero signals when no direct match is possible. Indeed, with conditional sampling (only nonzero signals are counted) the vacuum mode is easily filtered out in experimental sampling runs.

As initial condition, the optimization starts from nearly uncoupled waveguides, set as $J_{l,d}^{\text{(init)}}=0.01$, so that the coherent uncoupled input state (Eq.~\eqref{eq:init}) is contracted to a (near) zero-entanglement, separable circuit output state, well contained in the variational manifold set by the control parameter $\chi_\text{max}=500$. During the optimization, the gate couplings are constrained to be positive, with at most a full rotation from one waveguide to the other and back; $0\leq J_{l,d}\Delta t \leq \pi$). The couplings $J_{l,d}$ will gradually increase or damp out to zero across subsequent iterations of optimization. As additional constraint, we impose a regularization to avoid an unconstrained growth of the couplers between different iterations of optimization. Physically, this is justified by considerations related to chip fabrication; limiting the amount of nonzero couplers is beneficial. Indeed, couplers are sensitive to fabrication errors when two neighboring waveguides must approach each other at the correct separation (see simulation in Appendix \ref{app:MF2D}). On the other hand, this ensures numerically that the initial state remains well within the limits posed by the maximal MPS bond dimension, $\chi_\text{max}$. 

The Von Neumann entropy entanglement of bipartite chain divisions is directly accessible, thanks to the MPS being in right-canonical form, from the singular values stored in memory after two-mode gate contraction. This provides an adequate measure for regularizing $J_{l,d}$ in order to prevent unconstrained growth of the tensor bond dimensions $\chi_l$. To this end, we include in the FOM a measure for reducing von Neumann entropy,
\begin{equation}
\label{eq:LS}
    \mathcal{L}_S = \sum_{K=1}^{L} \big(S^{(K)}_{VN}\big)^2,
\end{equation}
with,
\begin{eqnarray}
\label{eq:S_ent}
\nonumber S^{(K)}_{VN} &=& -\text{tr}\big[\rho^{(l\leq K)} \ln{\rho^{(l\leq K)}}\big]\\ &=& -\sum_{k=0}^{\chi_K} (s^{(K)}_k)^2 \ln (s^{(K)}_k)^2,
\end{eqnarray}
with $L$ the number of waveguides (modes) and $\rho^{(l\leq K)}$ the reduced density matrix of modes left from waveguide $K$ ($l\leq K\leq L$). $S^{(l)}_{VN}$ is the bipartite entanglement entropy and $\chi_K$ the bond dimension (number of singular values), with $s^{(K)}_k$ the $k$-th left singular value of mode $K+1$. An extra (trivial) averaging over entanglement profiles must be done for the set of batch ensemble states in the Monte Carlo trajectory description.

Finally, the full FOM for optimizing state preparation is given as, 
\begin{equation}
\label{eq:Ltot}
    \mathcal{L}_\text{tot} = \lambda_\rho \mathcal{L}_\rho + \lambda_\text{S} \mathcal{L}_\text{S},
\end{equation}
with $\lambda_\rho$ the weight for the density matrix matching and $\lambda_\text{S}$ the weight for regularizing the state bipartite entanglement. We set $\lambda_\rho=1-\lambda_\text{S}$, as only the relative weight matters. For this use-case, the weights were picked empirically by hand, based on the system size $L$ and circuit depth $D$. As a control mechanism, the maximal bond dimension reached in the MPS was monitored, for a singular-value cutoff of $p^{(s)}_\text{min}=s^2_\text{min}=10^{-12}$, to restrict the growth in tensor bond dimensions under two-mode gate contraction and SVD. We targeted keeping, for all $l$, $\chi_l\lesssim500$ and the maximal value encountered under optimization was $\chi_l=325$.

\subsection{Case 1: Deep-circuit unitary cat-state generation}

\begin{figure*}[!ht]
    \centering
    \includegraphics[width=1\linewidth]{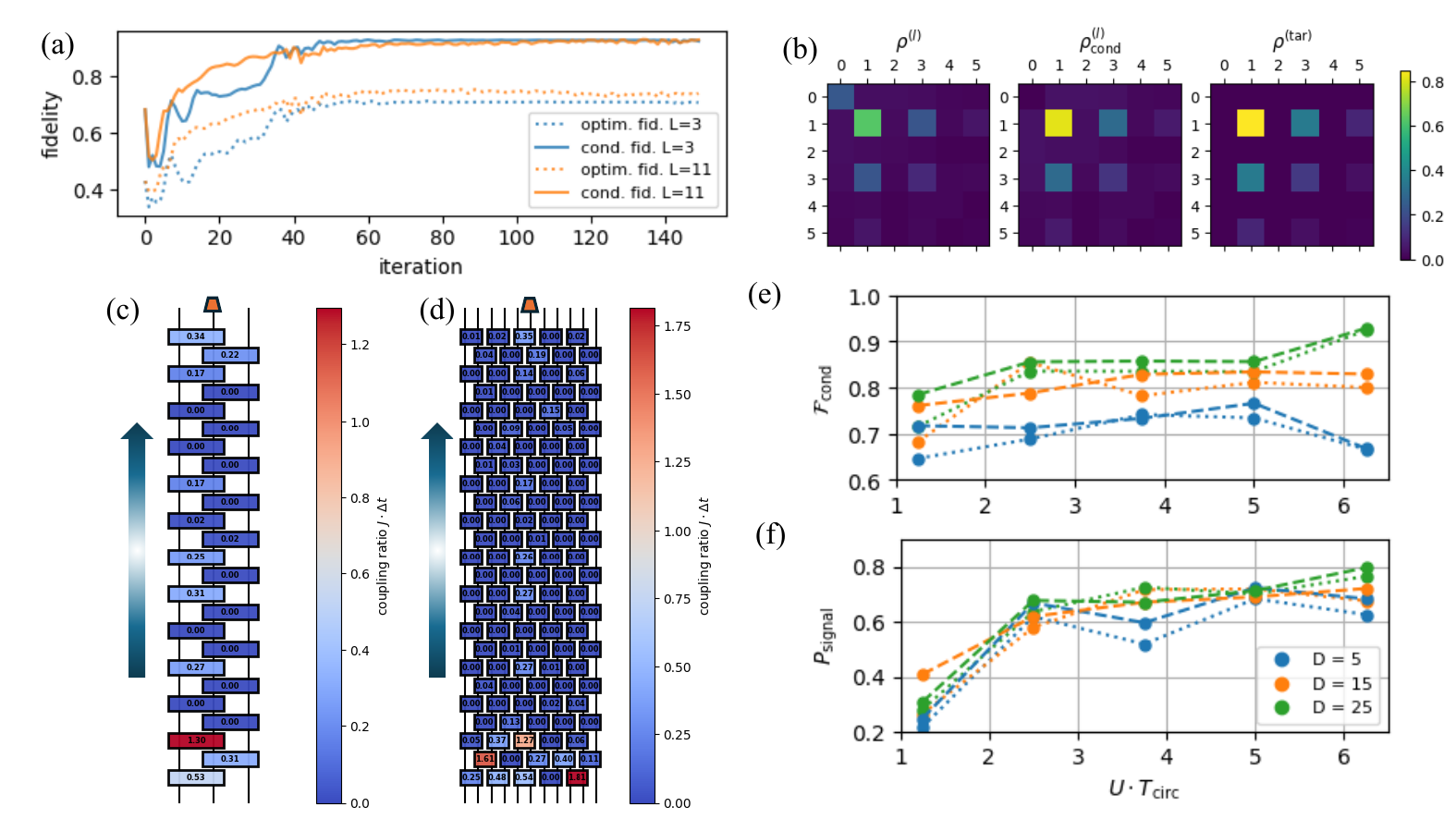}
    \caption{The results for cat state preparation. (a) The fidelity converges for $L=3$ (blue) and $L=11$ (orange), for both the optimization fidelity (dotted) and conditioned fidelity (full). (b) The density matrix obtained and compared with $\rho^\mathrm{(tar)}$ for $L=3$, both the unconditioned and conditioned. (c)-(d) The optimized circuits obtained, for $L=3$ and $L=11$, respectively, with circuit nonlinearity $U\cdot T_\text{circ}=6.25$. (e)-(f) The conditioned fidelity (e) and generation probability (f) as function of circuit accumulated nonlinearity, for $L=3$ (dotted) and $L=11$ (dashed lines), for different circuit depths $D$.}
    \label{fig:fig-cat}
\end{figure*}

We propose optimized qPICs to generate Schrödinger cat states, defined as,
\begin{equation}
\label{eq:cat}
    |\text{cat}_\pm\rangle = N_\pm\big(|\beta\rangle \pm |-\beta\rangle\big), \, N_\pm = \frac{1}{\sqrt{2(1 \mp e^{-2|\beta|^2})}},
\end{equation}
in the near-infrared regime, typical for excitons in GaAs samples, useful for fiber-based quantum networks. The even state $|\text{cat}_+\rangle$ only has even photon number occupation, while the odd state $|\text{cat}_-\rangle$ only has odd occupation. Cat states are crucial quantum resources in photonics due to their nonclassical superposition, which enhances quantum sensing, metrology, and information processing \cite{zheng2025quantum}, while showing robustness to phase noise \cite{mirrahimi2014dynamically}. The experimental preparation in optics is traditionally via photon number states and homodyne detection \cite{ourjoumtsev2007generation}.

In this use-case, we demonstrate the optimization of odd cat-state preparation with $\beta=1$ (Eq.~\eqref{eq:cat}) in the middle output of the circuit -- waveguide index $l=(L-1)/2$ for odd number of waveguides, or $l=L/2$ if even. For the optimization, the FOM from Eq.~\eqref{eq:Ltot} is used, with $\lambda_\text{S}=5\cdot 10^{-3}$. In the weighted trace distance from Eq.~\eqref{eq:Lrho}, a low weight is given to the vacuum mode $\rho^{(l)}_{00}$, by setting $\mathcal{M}_{00}$ a factor $9$ lower than other weights $\mathcal{M}_{(ij)\neq(00)}$. The maximal photon occupation is $N_\text{max}=5$, giving a physical dimension of $d=6$, and a learning rate of $0.1$ was used for the Adam optimizer. Coherent states of amplitude $\alpha=1$ are injected, so on average one photon per waveguide.

In Fig.~\ref{fig:fig-cat} the results are shown for cat-state generation as a function of (i) number of circuit waveguides, (ii) circuit depth and (iii) circuit accumulated nonlinearity $U\cdot T_\text{circ}$, with $T_\text{circ}$ the total time photons need to pass through the circuit. Points with the same $U\cdot T_\text{circ}$ (lying above each other) have the same circuit spatial extent, but that length is subdivided in a different number of layers $D$, thereby reducing the gate times $\Delta t$ proportionally. Therefore, higher $D$ offers more optimization ``flexibility'' for the couplers $J_{l,d}$ while preserving the circuit nonlinearity $U\cdot T_\text{circ}$.  For example, using a realistic baseline for the parameters, if the full circuit has $T_\text{circ}=2.5\,\text{mm}\cdot v_g^{-1}$, with $v_g$ the polaritonic group velocity, that means that for depth $D=25$ this would be a stack of layers of $100\,\mu\text{m}$ each, while $D=5$ would give $500\,\mu\text{m}$. To convert length scales to time units, we set $v_g$ a factor $2$ lower than the speed of light in GaAs, $0.5\,c/n_\text{GaAs}\approx 0.5\, \mu m \cdot\text{ps}^{-1}$, equivalent to $0.5$ exciton fraction -- see Sec.~\ref{sec:field}. This allows us to obtain dimensionful units for the photonic nonlinearity; e.g., $U\cdot T_\text{circ}=0.1$, leads to $U=5\cdot 10^{-4}\text{ps}^{-1}=0.34\,\mu \text{eV}/\hbar$, in turn corresponding to a 2D material nonlinearity $g = U\cdot S_\text{pulse} = 48\,\mu \text{eV}\cdot\mu \text{m}^2$, if the pulse associated surface is $141 \mu \text{m}^2$ -- see estimate in Appendix \ref{app:MF2D}, below Eq.~\eqref{eq:U_2D}.

In Fig.~\ref{fig:fig-cat}(a) the achieved state fidelity,
\begin{equation}
\label{eq:fidelity}
\mathcal{F(\rho)}=\bigg(\text{tr}\sqrt{\sqrt{\rho}\big(\rho^\mathrm{(tar)}\big)\sqrt{\rho}}\bigg)^2,
\end{equation}
is shown after $i$ iterations, for a narrow circuit of $L=3$ and a wide circuit with $L=11$ waveguides. We show $\rho^{(l)}$, having nonzero occupation of the vacuum mode due to the lower weight given in $\mathcal{M}_{00}$ (dotted lines), and the corresponding conditioned fidelity $\mathcal{F}_{cond}$, obtained after filtering out the vacuum occurrences (i.e., setting $\rho^{(l)}_{0i}=\rho^{(l)}_{i0}=0$) and renormalization (full lines). As expected, the conditioned fidelity is significantly higher. In Fig.~\ref{fig:fig-cat}(b) the obtained density matrix for $L=3$ is illustrated and compared with the target $\rho^\mathrm{(tar)}=\rho_{\text{cat}_-} = |\text{cat}_-\rangle\langle\text{cat}_-|$, for which a conditioned fidelity of $> 90\%$ is obtained for deep circuits of $D=25$. In Fig.~\ref{fig:fig-cat}(c)-(d), the optimized circuits are displayed, with the color codes representing the coupling strength of the gates. It is clear that, for the case $L=11$ (Fig.~\ref{fig:fig-cat}(d)), a large part of the gates converge to (near-) zero coupling (dark blue gate components), therefore only accumulating intra-waveguide nonlinearity. Indeed, the fidelity obtained for $L=3$ (Fig.~\ref{fig:fig-cat}(c)) is nearly identical, and we see that the gates on the right side in the middle, which couple the second and third waveguide, converge to zero coupling $J\Delta t=0$, while the left ones generate the optimal two-mode interferences. In conclusion, we observe that for single-mode odd cat state generation, fine-tuned interferences and accumulated Kerr-nonlinearity using three waveguides suffices to stabilize accurate odd cat-state photon statistics. In Fig.~\ref{fig:fig-cat}(e)-(f), the conditioned fidelity (Fig.~\ref{fig:fig-cat}(e)) and obtained probability $P_\text{signal}$ for cat-state generation (Fig.~\ref{fig:fig-cat}(f)) is shown as a function of accumulated circuit nonlinearity $U\cdot T_\text{circ}$, for different circuit depth (number of layers, color codes) and number of waveguides, $L=3$ (dotted) and $L=11$ (dashed).

Generally, the flexibility offered by circuits subdivided in larger number of layers $D$ leads to higher fidelities and probability of generation. In Fig.~\ref{fig:fig-cat}(f), we observe that, while the conditioned fidelity for $L=3$ and $L=11$ is similar, the probability of odd cat-state signal, $P_\text{signal} = 1-\rho_{00}$, is about 3\% higher for $L=11$ ($P_\text{signal}=0.797$) than for $L=3$ ($P_\text{signal}=0.765$) for the case of high nonlinearity, $U\cdot T_\text{circ}=6.25$. However, the statistical significance of this finding is, arguably, rather limited.

In Appendix \ref{app:num_cat_state}, we discuss in full detail, based on the full dataset of numerical simulations, the central role of the parameter $\lambda_S$  for stabilizing the multivariate non-convex unitary optimization problem for cat-state generation (see Eq. \eqref{eq:Ltot}). Too high values $\lambda_S$ will quell down any growth in inter-waveguide entanglement entropy, thus numerically limiting the initial search space for optimization, resulting in poor convergence values. Too low values, on the other hand, lead to an uncontrolled growth of $J_{l,d}$, which generates high entanglement and tensor bond dimensions. This results in (i) heavy numerics due to slow circuit and gradient evaluations and (ii) the drifting away from local optima of state fidelity under gradient-based approaches, due to the strong non-convexity of the optimization. In this regard, the parameter $\lambda_S$ controls the trade-off between Hilbert-space exploration and convergence of optimization convergence (exploitation) during the optimization of single-mode cat-state generation. We also present a full analysis of the runtimes and memory load encountered, and make a test comparison with exact-simulation methods to highlight the potential of the MPS representation.

Finally, in Appendix \ref{app:robustness}, we analyze in detail the circuit robustness towards small fabrication errors in the couplings $J_{l,d}$ and we explain that multiplicative Gaussian noise sampling provides a reliable first assessment, considering dominant misalignments encountered in fabrication. As a result, we find there is no discernible benefit for the robustness of $L=11$ circuits w.r.t. $L=3$. This brings us to the conclusion that it is the accumulated nonlinearity that is is of primary importance for odd cat-state generation. Next to that, larger layer division $D$ helps to approach better optimal sequences of interferences between linear inter-waveguide couplings and intra-waveguide nonlinear phase shifts. No immediate benefit is found for scaling up to qPICs beyond $L=3$ waveguides. Perhaps this changes if circuit robustness is incorporated as an explicit requirement in the FOM for state optimization \eqref{eq:Ltot}, as was studied, e.g., in Ref. \cite{kosmella2024noise} for neuromorphic optical computing. 

\subsection{Case 2: Noisy maximization of antibunched statistics and single-photon efficiency}
\label{subsec:single-photon}

\begin{figure*}[!ht]
    \centering
    \includegraphics[width=1\linewidth,trim={0 4cm 0 0},clip]{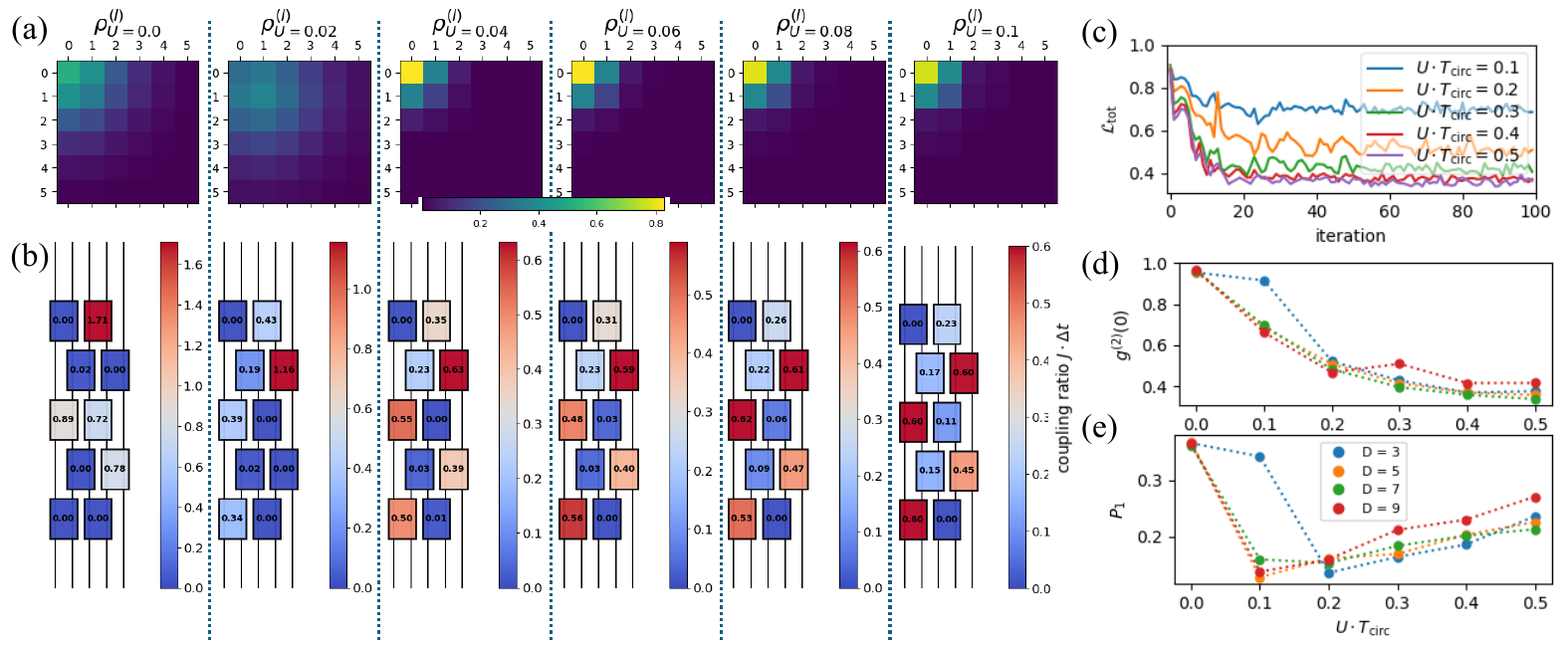}
    \caption{The results for noisy single-photon generation. (a)-(b) The obtained optimal density matrix (a) and corresponding optimal circuit (b), for different values of nonlinearity $U$. (c) The convergence of the FOM, for different values of $U$ at $D=5$. Due to the MC sampling ($N_s=400$), some noise remains after convergence. (d) the optimal value for $g^{(2)}(0)$ and (e) the probability of single-photon generation, for different circuit divisions $D$ (same color codes). Overall, higher $U\cdot T_\text{circ}$ leads to stronger antibunching and higher single-photon probability.}
    \label{fig:fig-sp}
\end{figure*}

Single-photon generation in the near-infrared regime is a key quantum resource for photonic quantum technologies and fiber-based communication. Such sources are generally realized via solid-state quantum emitters, defects in crystals, or nonlinear optical processes, enabling on-demand generation of pure single photons with high indistinguishability and brightness -- see Refs. \cite{chang2014quantum,jankowski2024ultrafast,liu2025single} for an overview. In this realm, strong antibunched photon statistics, characterized by the second-order correlation function $g^{(2)}(0):= \langle : n^2:\rangle / \langle n\rangle^2 < 1$, is a landmark for identifying nonclassical sources of light, ensuring the suppression of multi-photon events and high quantum fidelity. Generally, any measured value of $g^{(2)}(0)<1$, detectable within error bars, is considered a valid signature of ``nonclassicality'' of the photon statistics and a path towards photon blockade \cite{delteil2019towards}.

For the use-case of single photon generation in a noisy environment, we demonstrate the optimization scheme with the explicit inclusion of the stochastic Markovian loss processes. As previously outlined in Sec.~\ref{subsec:MC} (see Fig.~\ref{fig:MPS}), the MPS tensors $B_{l,ijk}$, with $l$ the waveguide mode index, are expanded in rank from $3$ to $4$, with an additional batch index $b$; $B^{(b)}_{l,ijk}$. This index serves for sampling the outcomes of the non-unitary stochastic gates $Q^{(b)}_l$ (Eq.~\eqref{eq:diss-gate}). The computational complexity increases for two reasons; (i) the added batch dimension and (ii) the contraction of non-unitarity single-mode gates $Q_{b,ij}$, which disrupts the canonical MPS form (Eq.~\eqref{eq:diss-gate}).

Our objective is to optimize the qPIC design under the practical constraints of fabrication techniques that are currently accessible, with the near-term goal of experimentally observing nonclassical photon statistics in waveguide-integrated polariton systems. To this end, we include as FOM the minimization of the second-order correlation function $g^{(2)}(0)$ for a chosen circuit output $l$:
\begin{equation}
\label{eq:Lg}
    \mathcal{L}_g := g^{(2)}_{ll}(0) 
    = \frac{\langle a^\dagger_l a^\dagger_l a_l a_l \rangle_{\text{out}}}
    {\langle a^\dagger_l a_l \rangle_{\text{out}}^2},
\end{equation}
where \(\langle O \rangle_{\text{out}} := \langle \psi_{\text{out}} | O | \psi_{\text{out}} \rangle\).
 In addition to this, we include the minimization of the single-photon density-matrix trace distance, as defined in Eq.~\eqref{eq:Lrho}, with $\rho^\mathrm{(tar)}_{11}=1$, $\rho^{(tar)}_{(ij)\neq(11)}=0$, leading to the FOM,
\begin{equation}
\label{eq:Lsp}
    \mathcal{L}_\text{tot} = \lambda_\rho\mathcal{L}_\rho + \lambda_g\mathcal{L}_g
\end{equation}
A larger weight was given to minimizing density-density correlations, $\lambda_g=0.9$ in comparison with the trace-norm single-photon matching, $\lambda_\rho=0.1$. Assuring $\lambda_\rho\neq 0$ secures that the mode $l$ does not converge to a value arbitrarily close to vacuum output for low nonlinearity $U$, as is generally seen in works on the \emph{unconventional photon blockade} -- see, e.g., Refs. \cite{lemonde2014antibunching,flayac2017unconventional}.

Similar as for cat state generation (Sec.~\ref{subsec:cat}), we consider a circuit architecture, now consisting of $L=5$ waveguides, subdivided in $D$ layers, with a circuit accumulated non-linearity $U\cdot T_\text{circ}$ and $T_\text{circ}$ the total circuit time, and the couplers are initially set to $J^\text{(init)}_{l,d}=0.01$. For this case, the entropy weight is largely irrelevant and set to zero. Indeed, a representation of the full Hilbert space of $L$ bosonic modes with truncated Fock space $\mathcal{N}_\text{max}$ and $L=5$ modes requires a MPS tensor bond dimension in the middle of the chain of $\chi_{L/2}=d^2=36$, with $d=\mathcal{N}_\text{max}+1=6$ the physical tensor dimension. Finally, similar as before, we set the weight for matching the vacuum mode, $\mathcal{M}_{00}$ a factor $9$ lower than other matrix elements in $\mathcal{L}_\rho$. Each optimization run was sampled with $\mathcal{N}_s=400$ stochastic sample states $b$, leading to a rank-4 MPS tensor representation with maximally $\mathcal{N}_s\cdot \chi_l^2d\leq400\times36\times6\times36\approx3\cdot 10^6$ tensor elements.

In Fig.~\ref{fig:fig-sp}, the results for noisy qPIC optimization for single-photon generation are shown, scanning the circuit depth $D$ and the accumulated nonlinearity $U\cdot T_\text{circ}$, with a constant loss rate of $\gamma\cdot T_\text{circ} = 0.5$ -- that is, on average $60\%$ of photons pass through circuit. To put it in numbers, if we set $T_\text{circ}=500 \, \mu \text{m} \cdot v_g^{-1}$, we scan gate lengths from $166\,\mu \text{m}$ ($D=3$) to $56\, \mu\text{m}$ ($D=9$), with a decay time $\tau=1/\gamma=5\,\text{mm}\cdot v_g^{-1}$. If we set the circuit nonlinearity $U\cdot T_\text{circ}=0.5$, the maximal value considered, the same values for circuit nonlinearity $\hbar U = 0.34\, \mu\text{eV}$ and material 2D nonlinearity $g=48\,\mu\text{eV}\cdot\mu\text{m}^2$ are obtained as estimated in Sec.~\ref{subsec:cat}.

We present in Fig.~\ref{fig:fig-sp} the obtained optimal density matrices (Fig.~\ref{fig:fig-sp}(a)) with the corresponding circuit configuration Fig.~\ref{fig:fig-sp}(b)), for increasing $U\cdot T_\text{circ}$, at $D=L=5$. For higher $U\cdot T_\text{circ}$, the number distribution grows more concentrated around $n=0$ and $n=1$ photons, suppressing photon number states of $n\geq 2$, as expected for achieving low values of $g^{(2)}_{ll}(0)$. Indeed, if exactly $\rho^{(l)}_{nn}=0$ holds for all $n\geq 2$, we get perfect antibunched statistics, $g^{(2)}_{ll}(0)=0$ -- see Eq.~\eqref{eq:Lg}.

In Fig.~\ref{fig:fig-sp}(c), the curves of convergence of the FOM $\mathcal{L}_\text{tot}$ from Eq.~\eqref{eq:Ltot} are given for different nonlinearities $U\cdot T_\text{circ}$ for circuit depth $D=5$. After an initial stage of convergence (roughly the first $20$ iterations), the optimal circuit shows statistical fluctuations in the semi-steady state it reaches in later iterations. Each iteration, a new batch of $\mathcal{N}_s=400$ Monte Carlo states was sampled, independently of the previous batch. In Fig.~\ref{fig:fig-sp}(d)-(e), the mean second-order correlations $g^{(2)}(0)$ and single-photon probability $P_1 = \rho^{(l)}_{11}$ are shown, respectively, for increasing accumulated nonlinearity $U\cdot T_\text{circ}$ and varying circuit division $D$. The final $20$ iteration were averaged out to obtain the data points. The second-order correlations $g^{(2)}(0)$ converges to similar values for the second-order correlation function at high nonlinearity, $g^{(2)}(0)\approx 0.4$, largely independent of circuit division depth $D$. From this we conclude that the circuit nonlinearity $U\cdot T_\text{circ}$ is the main factor for generating antibunched photon statistics and, to a much lesser extent, the flexibility offered by varying the couplers $J_{l,d}$, as quantified by the circuit depth $D$. When analyzing the single-photon output probability $P_1=\rho_{11}$ (see Fig.~\ref{fig:fig-sp}(e)), the probability of single-photon generation increases roughly linearly with nonlinearity $U\cdot T_\text{circ}$. Furthermore, a small increase ($\sim3\%$) is seen in the single-photon output probability for high-flexibility $D=9$. Though, it should be noted that at $D=9$ also $g^{(2)}(0)$ is slightly higher at higher nonlinearities $U\cdot T_\text{circ}\geq 0.5$. Therefore, if given more flexibility, the FOM for single-photon generation, formulated in Eq.~\eqref{eq:Lps}, gives preference to slightly higher probability of single-photon generation, at the cost of lower antibunching -- for the initial condition $J_\text{init}$ and weights $\lambda_g$ and $\lambda_\rho$ considered here. 

%The combined FOM $\mathcal{L}_\text{tot}$ converges to similar values for $U\cdot T_\text{circ}=0.4$ or $U\cdot T_\text{circ}=0.5$.

In summary, qPICs with $L=5$ waveguides were investigated for noisy single-photon generation using stochastic MC sampling methods. The focus was on a regime reachable with current experiments and fabrication methods -- see, e.g. Ref.~\cite{suarez2021enhancement}. qPICs of total spatial extent $\sim500\,\mu\text{m}$ were considered as an experimental baseline and it was observed that the circuit accumulated nonlinearity $U\cdot T_\text{circ}$ is of primary importance, much more than the flexibility in varying the gate couplings offered by an increased layer subdivision $D$. The observations were generated with $N_s=400$ sample states to illustrate the effect of the noise in the statistical sampling. When $\mathcal{N}_s$ is increased further, the standard deviation reached after the initial convergence is expected to decrease as $1/\sqrt{\mathcal{N}_s}$, as is the default for stochastic sampling~\cite{daley2014quantum}.

%% file: 4-metrology/metrology.tex
\section{Optimal readout for quantum sensing}
\label{sec:metrology}

Quantum metrology and sensing~\cite{giovannetti2011advances, degen2017quantum, pirandola2018advances, polino2020photonic,khan2025quantum} has numerous applications in a wide range of fields; among which, global positioning~\cite{kantsepolsky2023exploring}, gravitational-wave detection~\cite{tse2019quantum}, biomedical applications~\cite{aslam2023quantum} or chemical detection~\cite{kosterev2002chemical}. For quantum sensing, the purpose is to construct a quantum state, on which a small perturbation can be optimally detected with a positive operator-valued measure (POVM). Harnessing multipartite quantum statistics, bounds from classical detection statistics may be broken, thereby possibly reaching the Heisenberg quantum limit when an optimal measurement is performed on the quantum state. This is characterized by the classical vs. the quantum Cramer-Rao bound~\cite{yu2022quantum}. The high complexity of (i) state initialization, (ii) signal detection and (iii) measurement analysis for optimal quantum sensing has led to general optimization schemes \cite{yuan2022optimal}, possibly including Bayesian  optimization \cite{fiderer2021neural} or machine learning \cite{fallani2022learning}, also in optical setups \cite{pezze2023machine,munoz2024photonic}.

\begin{figure*}[!ht]
    \centering
    \includegraphics[width=1\linewidth, trim={0 4.5cm 0 0},clip]{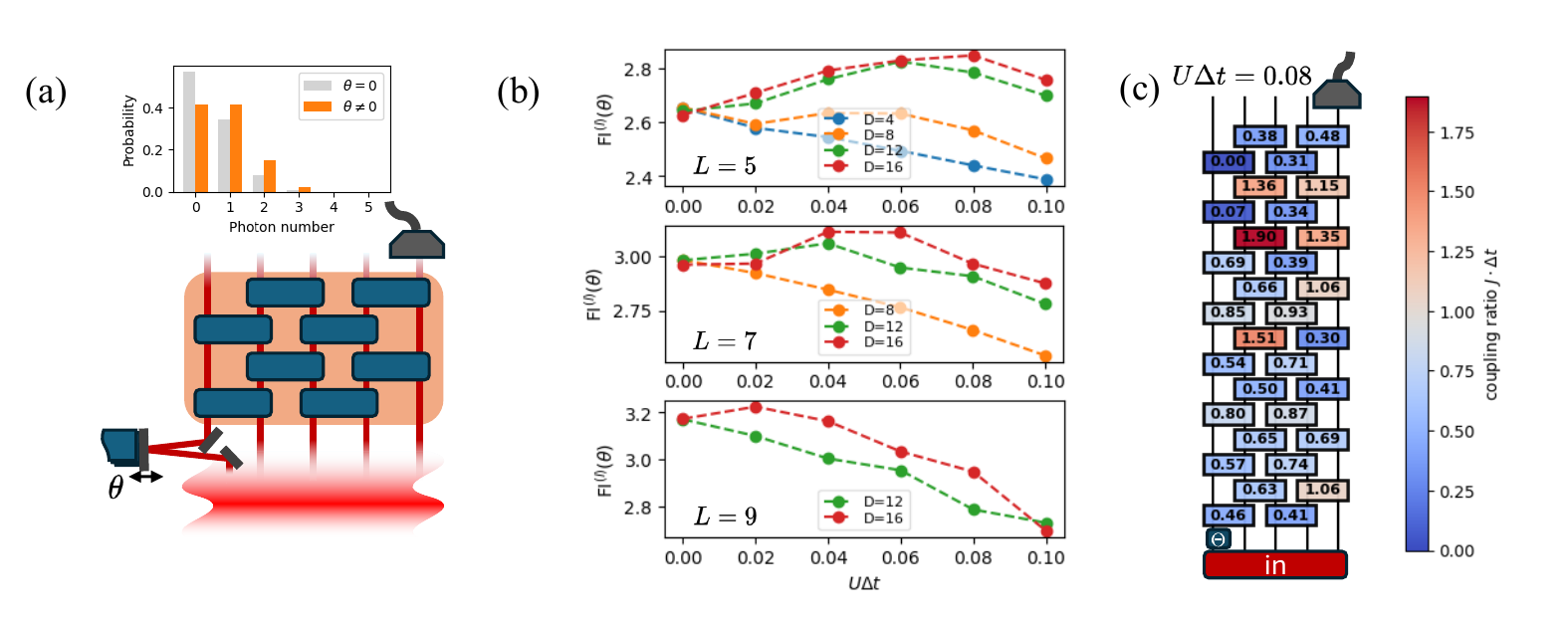}
    \caption{Circuit optimization applied to optimal readout for phase sensing. (a) an illustration of the setup, one input receives a phase shift from an object displacement, from one output the number distribution is used for detection. (b) the results for optimizing the measurement Fisher information for different waveguides, $L=5$ (top), $L=7$ (middle) and $L=9$ (bottom) and different circuit depths (color codes). (c) a visualization of the optimal circuit obtained for $L=5$ and $D=16$, for $U\Delta t=0.08$,  a phase shift is given to the left input, while the readout is in the right output.}
    \label{fig:sensing}
\end{figure*}

In this analysis, we focus on the task of signal detection, generated with an optimal qPIC for the sensing of a small incoming phase perturbation in the regime of low-photon occupation, using the quantum statistics generated by a nonlinear circuit gates coupling the integrated waveguides -- see Fig.~\ref{fig:sensing}. Same as before, a set of uncoupled coherent in-phase signals (see Eq.~\eqref{eq:init}) enters the qPIC, but, for this case, one of the incoming signals is given a small phase shift to be sensed, possibly caused by a slight displacement of a reflective object (as illustrated in Fig.~\ref{fig:sensing}(a)), a perturbance in material refractive index, or a small variation in exciton Stark shift (see, e.g., Ref. \cite{liran2024electrically}). The waveguides $1\dots L-1$ receive the same coherent in-phase laser input $|\alpha\rangle$, $\alpha_{l \geq 1}=1$, while the first waveguide is slightly perturbed, with $\alpha_{l=0}=e^{i\theta}$ and $\theta$ infinitesimally small.  We develop an optimization scheme for maximizing the sensitivity towards the phase $\theta$ of the first input mode $l=0$ in the circuit output signal of the last output mode, $l=L-1$. The regime of low-photon occupation is explicitly considered, with the injection of $|\alpha|^2=1$ photon on average per waveguide mode $l$. This is crucial in settings where high photon flux could damage sensitive samples, such as biological specimens, or where low-energy probes are needed to avoid perturbation.

\subsection{The figure of merit for optimizing phase sensing}
We consider maximizing the classical Fisher information (FI) of the photon number distribution of one single circuit output mode $l$, 
\begin{eqnarray}
\label{eq:FI}
    \text{FI}^{(l)}(\theta)\Big|_{\theta=0} &=& \sum_n P_n^{(l)}(0)\, \Big( \partial_\theta \log P_n^{(l)}(\theta)\big|_{\theta=0} \Big)^2
    %\\&=& \sum_n P_n^{(l)}(0) \times \lim_{\theta\rightarrow 0} \frac{1}{\theta^2}\big(\log{P_n^{(l)}(0}) -\log{P_n^{(l)}(\theta})\big)^2,
\end{eqnarray}
with $P_n^{(l)}(\theta) = \rho^{(l)}_{nn}(\theta)$, the photon number distribution of the final output $l\equiv L-1$, conditioned on the input phase difference $\theta$ ($\theta=0$ if unperturbed). The reduced density matrix of the waveguide mode, defined as $\rho^{(l)}\big(\{J_{l,d}\}\big) = \text{tr}_{l'\neq l}\Big[\big|\psi_\text{out}\big(\{J_{l,d}\}\big)\big\rangle\Big]$, is obtained from the tensor contraction of the qPIC output state -- see Fig.~\ref{fig:MPS}(c).

The classical Fisher information provides an immediate bound on the variance for metrology of the variable $\theta$ from these measurement statistics \cite{yu2022quantum},
\begin{equation}
    \label{eq:FIvar}
    \big(\Delta \theta\big)^2 \geq \frac{1}{m\text{FI}^{(l)}(\theta)} \geq \frac{1}{m\text{QFI}(\theta)},
\end{equation}
with the variance $\Delta \theta^2=E\Big[\big(\theta - \mu_\theta\big)^2\Big]$ and $m$ the number of independent statistical measurement samples -- the number of pulses detected at the circuit output, in this case. $\text{QFI}(\theta)$ is the quantum Fisher information, which bounds the classical $\text{FI}(\theta)$ that can be extracted from a quantum state with a POVM. The Heisenberg limit for detection is saturated when the second bound in Eq.~\eqref{eq:FIvar} is satisfied, i.e., $\text{FI}^{(l)}(\theta)\equiv\text{QFI}(\theta)$. For the case we are considering, $\text{QFI}(\theta)$ can be evaluated analytically from coherent state overlap and gives $\text{QFI}(\theta)=4|\alpha|^2$ -- see Appendix \ref{app:QFIcoherent}.

Here, we want to maximize $\text{FI}^{(l)}(\theta)\big|_{\theta=0}$, for the purpose of minimizing the variance on detecting a small phase shift from $\theta=0$, $(\Delta\theta)^2$, while approaching the Heisenberg limit as close as possible. In other words, the qPIC is optimized to emulate a Hermitian operator for measurement that maximizes the extraction of Fisher information $\text{FI}^{(l)}(\theta)$. For this, we formulate the FOM for optimizing phase-sensing qPIC readout as,
\begin{equation}
    \label{eq:Lps}
    \mathcal{L} = \lambda_\text{ps} \mathcal{L}_\text{ps} + \lambda_
\text{S}\mathcal{L}_\text{S},
\end{equation}
with $\mathcal{L}_\text{ps} = - \text{FI}^{(l)}$. Same as for state generation, the coupling rates $J_{l,d}$ are regularized by reducing the state bipartite entanglement, using the objective $\mathcal{L}_S$ -- see definition in Eq.~\eqref{eq:LS}.

\subsection{Optimal qPIC for phase-sensing readout}

In Fig.~\ref{fig:sensing}, we present the results obtained for optimizing the unitary qPIC for the task of quantum phase sensing, without considering dissipation. In Fig.~\ref{fig:sensing}(a), the setup is illustrated, the left coherent input (index $l=0$) is given a small phase shift and the qPIC and the distinguishability towards the number distribution of the last output (index $l=L-1$) is maximized, using the FOM from Eq.~\eqref{eq:Lps}. This way, the effect of multimode interferences is maximized as a starting condition, by the need for the signal to pass the circuit from left to right. The experimental signature for this is characterized by the classical Fisher information of one output's photon number distribution -- see Eq.~\eqref{eq:FI}. The goal is to approach the quantum Heisenberg limit as close as possible, variationally within the qPIC architecture, as set by the quantum Fisher information (QFI). As derived in Appendix~\ref{app:QFIcoherent}, with $\alpha=1$, this sets a bound of $\text{QFI}=4$, so that $\mathcal{L}_\text{ps}\geq-4$. The results are shown in Fig.~\ref{fig:sensing}(b)-(c), for different circuit size $L\in\{5,7,9\}$ (up-down) and for different depth $D\in\{4,8,12,16\}$, with $|\alpha|=1$ across all inputs. Additionally, the initial condition for the couplers was set relatively high to start from large interferences, $J\Delta t=0.3$ -- remember, $J\Delta t=\frac{\pi}{4}\approx0.78$ is a 50:50 coupler. Note that $D\geq L-1$ must hold for the phase-shifted signal to be connected to the output measurement when trespassing the circuit.

We find that, for sufficiently deep circuits, the optimal Fisher information is obtained at nonzero nonlinearities, $U\Delta t>0$. This is clearly seen for $L=5$ and $L=7$, which have an optimum at depths $D=12$ or $D=16$ for higher, nonzero gate nonlinearities. Remarkably, this makes us conclude that, by gathering number outcome statistics in the low-photon occupation regime, it is not solely linear photonic interferences that realize maximal phase-shift sensitivity (as would be done, e.g., in standard homodyne detection schemes) but that higher circuit nonlinearity $U\cdot T_\text{circ}>0$, with $T_\text{circ}=D\Delta t$, can surpass the classical limit of linear photonics at $U=0$, where all circuit depths $D$ coincide. When $U\Delta t$ becomes to high, effects of blockade, caused by the nonlinearity-induced blueshift, become dominant, as such reducing the qPIC phase sensitivity -- see, e.g., the experimental results in~\cite{liran2024electrically}.

In Appendix~\ref{app:gaussian_metrology}, we explain in more detail that the optimal circuit sensitivity in the regime of low pulse photon occupancy can be understood, to a limited extent, from an analysis of Gaussian FI (see, e.g. Ref. \cite{kay1993fundamentals}). It is shown that the Gaussian FI, given by $\frac{1}{\sigma^2}\big( \partial_\theta \mu\big)^2$, with mean $\mu=\langle n_l\rangle$ and variance $\sigma^2=\langle n_l^2\rangle - \mu^2$, accounts for the large part of the optimal sensitivity found (it is even exact for the case $U=0$). Nevertheless, the genuinely non-Gaussian part of the photon number statistics is crucial to explain the actual advantage seen for deep circuits with a nonzero gate nonlinearity $U\Delta t$. 

Finally, in Fig.~\ref{fig:sensing}(c), the design of an optimal circuit is illustrated for the case $L=5$, $D=16$ and $U\Delta t=0.08$, the optimum of the red curve in Fig.~\ref{fig:sensing}(b). As can be seen, there is no immediate discernible pattern found in the optimal coupler values $J^{(opt)}_{l,d}$, which are, overall, rather high and close to maximal coupling $\frac{\pi}{4}\approx 0.78$.

From experience, We noticed that the convergence of the FOM (Eq.~\eqref{eq:Lps}) depends crucially on the initial condition given to the couplers. When $J^\text{(init)}$ (set uniform across circuit) is too low, the gradient vanishes due to the exponentially damped connection between the phase-shifted input ($l=0$) and analyzed output ($l=L-1$), resulting an exponentially slow convergence. This is an issue that could be understood from the more general perspective of \emph{barren plateaus} in variational quantum computing \cite{larocca2025barren}.

In Appendix~\ref{app:homodyne-sensing}, we study the FI associated with standard homodyne detection schemes, using one linear beam splitter. For the homodyne phase detection of a signal $\alpha_s$ using a local oscillator $\alpha_\text{LO}$, with $|\alpha_s|=|\alpha_\text{LO}|=1$, we derive that $\text{FI}(\theta)=1$, with an additional gain of a factor $2$ when both outputs are observed simultaneously. In this regime of low photon occupation, we demonstrate in Fig.~\ref{fig:sensing}(b) that a significant benefit can be achieved by using a qPIC with multiple waveguides (which serve as local oscillators) and a stacked pattern of nonlinear gate interferences; this leads to values of FI well above the values reachable with homodyne detection. 

To conclude this part, we presented a procedure for optimizing the qPIC phase readout capability in the limit of low photon occupation by analyzing the photon number distribution $P_n$ of one output signal. It was found that, while multiple linear photonic interferences provide a significant benefit over standard homodyne detection for the extraction of FI, adding modest circuit gate nonlinearities $0<U\cdot \Delta t \lesssim 0.1$ can provide an additional improvement. This is not fully understood by solely considering the Gaussian statistics of the output light, but requires a perception of the full quantum photon number statistics.

In the future, more generally, both a stage of state preparation and of state readout can be optimized for the maximizing the qPIC phase-sensing potential, as explained in Ref. \cite{giovannetti2011advances}. Recently, this was studied for two-mode photonic circuits in Ref. \cite{munoz2024photonic}, using a Kerr-type nonlinearity, and this can be generalized to larger qPICs.

%% file: 5-conclusions-outlook/conclusions_outlook.tex
\section{Conclusions and outlook}
% We presented a novel \mvrc{better wording...} method for optimizing photonic integrated quantum circuits using tensor-network quantum circuit description and gradient-based descent methods. 
We developed a method for qPIC design optimization in the low-photon occupation regime, enabling the capture of quantum statistics throughout the optimization process. This was achieved by deploying a differentiable tensor-network representation of the bosonic field of a photon pulse in the qPIC waveguides, thus allowing for (i) efficient unitary circuit contraction with the possibility to include non-unitary loss channels for the evaluation of the figure of merit and (ii) gradient tracking for optimizing the qPIC design. 

Starting from classical field equations, we established a mapping to a design-level description of photonic quantum circuits, thereby providing a practical framework for the optimization of qPICs. Photonic losses can be incorporated during the optimization, using Monte Carlo sampling methods encoded in dissipative single-mode loss gates which preserve the differentiable nature of the tensor contraction. 

The method was put in practice for three use-cases: in the realm of quantum state preparation, unitary qPIC odd cat-state generation and noisy single-photon generation were considered, and for quantum sensing, the optimal qPIC sensitivity was achieved by reading out the photon number distribution. For cat-state generation, we found that the circuit-accumulated nonlinearity, combined with sufficient iterations of three-waveguide nonlinear interference suffices for high-fidelity and high-probability success outcomes. 

For single-photon generation, the dissipation induced by strongly coupling the light to the excitonic degree of freedom was explicitly incorporated in the optimization procedure, using Monte Carlo sampling techniques, to maximize the generation of strong antibunched photon statistics statistics, while simultaneously maximizing output intensity. We concluded that, while the degree of antibunching primarily depends on the circuit-accumulated nonlinearity, subdivisions in more circuit layers lead to slightly higher probabilities of single-photon generation. 

Finally, we demonstrated the case of optimal readout of quantum sensing by maximizing the circuit sensitivity towards a small phase shift of the incoming coherent light. We found that deep qPICs can potentially offer a significant benefit for the Fisher information extracted by analyzing the output photon number distribution. The multimode qPIC photonic interferences result in a significant improvement over standard homodyne methods. Moreover, a nonzero photonic nonlinearity leads to genuine non-Gaussian photon statistics of the output signal, which, in turn, can effectuate an additional benefit over linear multimode processing.

The proposed optimization scheme for qPICs can be extended to enhance several other contemporary use-cases in quantum processing. To support broader adoption, we have open-sourced the developed code in \cite{coderepo}. As a first continuation, quantum phase sensing and metrology can be further improved and finetuned. As of now, we presented an optimal phase-sensing readout scheme for coherent light, but, more generally, the two stages of state initialization and state readout can be optimized for maximizing the multimode phase sensitivity \cite{pirandola2018advances,munoz2024photonic}. The simultaneous optimization of both may lead to an enhanced scheme, in which the advantage of the multimode nature of the quantum light is maximally profited from when extracting the Fisher information from qPIC readout, thereby breaching scaling limits from classical signal detection. Second, while we showed a brief analysis of the circuit robustness for cat-state generation (see Appendix \ref{app:robustness}), this can be expanded in future works and should be explicitly considered for any real-world use case. To this end, in classical integrated photonics, several methods exist to perform noise-aware optimization of a certain processing tasks -- see, e.g., Ref. \cite{gu2022,kosmella2024noise}. 

In addition to that, also the sparsity of the circuit can be optimized further to limit the impact of fabrication errors in the final detection. Moreover, we now only analyzed the possibilities of coherent light as input, with the conditioning of nonzero output signal, but this can be further expanded. For example, considering single-photon input in one or more waveguides, or conditioning output state-generation of one waveguide on the detection of single (or multiple) photon clicks in another, provides valuable resources for augmenting the quantum nature of the photonic output signal.

In terms of simulation methods, we proposed an optimization scheme based on tensor-network contraction, valid for photonic quantum states in the low occupation regime. The limited set of tensor operations is highly convenient and meory-efficient for tracking the gradients towards circuit parameters. Yet, this imposes sharp limits on (i) the photonic Fock space (mode occupation) and (ii) the bipartite entanglement that can be reliably encapsulated. In that sense, for larger photonic mode occupation and intermode entanglement, considering continuous-variable simulation methods could be beneficial for developing the circuit optimization. For example, the positive-P representation \cite{gardiner2004quantum}, recently used for describing quantum optical neural networks \cite{swierczewski2025phase}, does not impose a sharp cutoff on photon occupation, but it faces difficulties, due to the multiplicative noise sampling, in the regime of high occupation, strong nonlinearity and weak dissipation \cite{deuar2021fully}. These challenges must be compared more closely with the ones from the tensor description of the photonic field. 

Furthermore, in the weakly interacting regime, approximate variational simulation methods may suffice for developing the qPIC optimization. In that realm, deterministic variational methods in the Gaussian manifold of quantum states (also known as Hartree-Fock-Bogoliubov \cite{tikhonenkov2007quantum}) can be considered, possibly incorporating higher-order non-Gaussian cumulants \cite{van2017spontaneous}, or approximate stochastic methods, in particular the truncated Wigner approach \cite{gardiner2004quantum,deuar2021fully}. With those methods, much larger circuits can be simulated, using more bosonic modes and gate operations, without inflicting stringent constraints on photonic occupation or multimode entanglement statistics. Nevertheless, this comes at the cost of capturing only a limited and well-contained part of the inherent photon statistics that designate the photonic quantum nature of light. 

To finish, while we illuminated GaAs-based planar samples as the baseline for the present work, a plethora of different materials and coupling mechanisms can be explored for integrating the photonic architectures. As we briefly mentioned in the introduction, heterogeneous SiN-GaAs samples are gaining broad attention, thanks to the compact, broad transparency window and well-controlled linear processing characteristics of SiN structures. This can be addressed with a modest reparameterization of the qPICs for optimization (see Eq. \ref{eq:V2_SiN}). Recently, also 2D materials are gaining ample interest for devising strong coupling to light. Furthermore, vibrational couplings (Raman modes), plasmonics or coupled chains of cavity systems are amongst the possibilities for further exploration and characterization.

%% file: appendix/appendix.tex
% \begin{widetext}
% \begin{center}
%     \Huge\textbf{Appendix}
% \end{center}
% \end{widetext}

\section{Deriving the gate dynamics of the qPIC from full 2D field simulations}
\label{app:MF2D}
We work out a blueprint for the gate composition achievable in GaAs-based photonic integrated circuit using simulations of the full 2D polaritonic field, defined in the plane in which the waveguide circuit topology is etched. A short introduction is given to describe the method.

\subsection{The 2D field equations}
The 2D Hamiltonian for the chip is expressed in terms of the bosonic photon field operators $\hat{\psi}_\text{ph}(\vec{r})$ and $\hat{\psi}_\text{ex}(\vec{r})$ for the photons and excitons, respectively,
\begin{equation}
\label{eq:2dham}
    \hat{\mathcal{H}} = \hat{\mathcal{H}}_\text{ph-ex} + \hat{\mathcal{H}}_\text{int} + \hat{\mathcal{H}}_\text{pot}.
\end{equation}
The first term gives the photon and exciton dispersion and the photon-exciton coupling,
\begin{widetext}
\begin{equation}
\label{eq:H-ph-ex}
    \hat{\mathcal{H}}_\text{ph-ex} = \int d^2\vec{k} \;\Big[\hbar\omega_\text{ph}(\vec{k})\hat{a}_\text{ph}^\dagger(\vec{k}) \hat{a}_\text{ph}(\vec{k)} + \epsilon(\vec{k}) \hat{a}_\text{ex}^\dagger(\vec{k}) \hat{a}_\text{ex}(\vec{k)}\Big]
    + \frac{\hbar \Omega_R}{2}\int d^2\vec{k} \Big[ \hat{a}_\text{ph}^\dagger(\vec{k}) \hat{a}_\text{ex}(\vec{k}) +\hat{a}_\text{ex}^\dagger(\vec{k}) \hat{a}_\text{ph}(\vec{k})\Big].
\end{equation}
\end{widetext}
Here, $\hat{\psi}_\text{ph,ex}(\vec{r}) = 1/\sqrt{V} \int d^2\vec{k}\; e^{i\vec{r}\cdot \vec{k}}\,\hat{a}_\text{ph,ex}(\vec{k})$ is the Fourier transform of the photon (ph) or exciton (ex) field. The first term in \eqref{eq:H-ph-ex} represents the bare in-plane photon dispersion $\omega_\text{ph}(\vec{k})$ and exciton dispersion $\epsilon(\vec{k})$. We assume the photon dispersion to be effectively linear, that is $\omega_\text{ph}(\vec{k})=c_\text{eff}|k|$, with $c_\text{eff}=c/n$ the speed of light in the material ($c$ the speed of light in vacuum and $n$ the relative refractive index of the material), while the exciton dispersion is constant, $\epsilon(\vec{k})=\hbar \omega_\text{ex}$, as it represents (largely) immobile massive excitons, with negligible dispersion as compared to relevant photonic energies. The second term in Eq.~\eqref{eq:H-ph-ex} describes the exciton-photon coupling in the strong coupling regime, given by the Rabi frequency $\Omega_R$ \cite{carusotto2013quantum}. For maximal in-plane coupling, we assume the photonic field to be in transverse electric (TE) zero eigenmode \cite{shapochkin2018polarization}, so that only one polarization component of the exciton and photon field is considered.

The second term in Eq.~\eqref{eq:2dham} describes the local exciton-exciton interaction,
\begin{equation}
\label{eq:H-int}
    \hat{\mathcal{H}}_\text{int} = \frac{g_{2D}}{2}\int d^2 \vec{r}\; \hat{\psi}^\dagger_\text{ex}(\vec{r})\,\hat{\psi}^\dagger_\text{ex}(\vec{r})\, \hat{\psi}_\text{ex}(\vec{r})\, \hat{\psi}_\text{ex}(\vec{r})\,,
\end{equation}
with $g_{2D}$ the 2D interaction constant of the contact interaction excitons experience in the plane. Since uniform polarization of the light is considered (indistinguishable photons), a repulsive fermionic exchange interaction is dominant, with $g_{2D}>0$ \cite{vladimirova2010polariton}. When applying an external electric field, the induced exciton dipole-dipole interaction amplifies the effective interaction strength, possibly several orders of magnitude \cite{tsintzos2018electrical,suarez2021enhancement,liran2024electrically}.

Finally, the last term in \eqref{eq:2dham} is the potential experienced by the photons for the transversal confinement in the etched waveguide configuration,
\begin{equation}
    \hat{\mathcal{H}}_\text{pot} = \int d^2 \vec{r}\; V_\text{ph}(\vec{r}) \hat{\psi}^\dagger_\text{ph}(\vec{r}) \hat{\psi}_\text{ph}(\vec{r}).
\end{equation}
The photonic potential $V_\text{ph}(\vec{r})$ is engineered by etching the 2D GaAs sample, which locally modifies the in-plane effective 2D refractive index experienced by the light (see, e.g., Ref. \cite{suarez2021enhancement}). This can be understood immediately from the wave equation of the classical photonic field, $\psi_\text{ph}=\langle \hat{\psi}_\text{ph} \rangle$
\begin{equation}
    \nabla^2\psi_\text{ph} - \frac{n^2(\vec{r})}{c^2}\partial_t^2 \psi_\text{ph} = 0,
\end{equation}
from which the Helmholtz equation follows for monochromatic light, $\psi_\text{ph}(\vec{r}, t) = \phi(\vec{r})e^{i\omega t}$,
\begin{equation}
\nabla^2 \phi + k^2_0 n^2(\vec{r}) \phi = 0,
\end{equation}
with $k_0=2\pi/\lambda=\omega/c$, with $\lambda$ the wavelength. Comparison with the Schr\"odinger equation and recalibration of energy tells us that,
\begin{eqnarray}
    V(\vec{r}) = k^2_0\big(n_0^2 - n^2(\vec{r})\big) \approx - 2 k_0^2 n_0 \,\Delta n(\vec{r}),
\end{eqnarray}
which holds for $|n_0 - n(\vec{r})| \ll n_0$, with $n_0$ the material refractive index and $n(\vec{r}) = n_0 + \Delta n(\vec{r})$ the etched index landscape.

The potential to shape the 2D photonic potential landscape by imprinting waveguide designs on the chip is a core component of the research presented in the main text. Identifying and quantifying the effective circuit-gate coupling constant $J$ and nonlinearity $U$ of a gate design under some form of optical excitation will be explored in what follows.

\subsection{2D Gross-Pitaevskii gate simulation}
\label{subsec:2DGP}
The photon-exciton coupling Hamiltonian, $H_\text{ph-ex}$ from Eq.~\eqref{eq:H-ph-ex}, is diagonalized in momentum space, which gives the upper (UP) and lower polariton (LP) dispersion \cite{keeling2007collective, carusotto2013quantum},
\begin{widetext}
    \begin{equation}
    \label{eq:dispersion}
        \omega_\text{UP,LP}(\vec{k}) = \frac{1}{2}\left(\big(\omega_\text{ph}(\vec{k}) + \omega_\text{ex}\big) \pm \sqrt{\Omega_R^2 + \big(\omega_\text{ph}(\vec{k}) - \omega_\text{ex}\big)^2}\right),
    \end{equation}
    \end{widetext}
    with corresponding eigenmodes, 
    \begin{equation}
    \label{eq:ex_ph_mixing_LP_UP}
        \left(\begin{array}{cc}
             \hat{a}_\text{ph}(\vec{k})  \\
             \hat{a}_\text{ex}(\vec{k})
        \end{array}\right) = 
        \left(\begin{array}{cc}
             \cos\theta_\vec{k} & -\sin\theta_\vec{k}  \\
              \sin\theta_\vec{k} & \cos\theta_\vec{k}
        \end{array}\right)
                \left(\begin{array}{cc}
             \hat{a}_\text{UP}(\vec{k})  \\
             \hat{a}_\text{LP}(\vec{k})
        \end{array}\right),
    \end{equation}
    
    and $\tan \theta_\vec{k} = \frac{\Omega_R}{\omega_\text{LP}(\vec{k}) - \omega_\text{ph}(\vec{k})}$.
    In what follows, we assume the incident light is in close resonance with the LP branch, so that the excitation of the UP branch can be neglected. We restrict to the bosonic operators $\hat{a}:=\hat{a}_\text{LP}$, with a rescaled lower-polariton potential with $V:=V_\text{pol} = \sin^2\theta_\vec{k} V_\text{ph}$ and a LP interaction $g:=g_\text{LP,2D} = \cos^4\theta_\vec{k}\,g_\text{2D}$, with $u^2_\vec{k}=\cos^2\theta_k$ the excitonic and $v^2_\vec{k}=\sin^2\theta_k$ photonic fraction \cite{carusotto2013quantum}.

Following this, we find an effective model that describes the dynamics of the LP field $\psi(\vec{r},t):=\langle \psi_\text{LP}(\vec{r},t)\rangle $ on the level of mean-field dynamics, 
\begin{widetext}
\begin{equation}
\label{eq:GPE2D}
    i\partial_t\psi(\vec{r},t) = -i\gamma(\vec{r})\, \psi(\vec{r},t) + \mathcal{E}\big[\psi(\vec{r},t) \big] + g\big|\psi(\vec{r},t)\big|^2 \psi(\vec{r},t) + V(\vec{r}) \psi(\vec{r},t).
\end{equation}
\end{widetext}
Here, in general, a space-dependent polaritonic loss rate $\gamma(\vec{r})$ is added.

The dispersion experienced by the photons is approximated. The light is injected along $z$-direction, at narrow linewidth and in close resonance with the exciton level. The direction $x$, transverse to the propagation, is where the waveguide confinement takes place, with a length scale of the same order as the wavelength of the light. Therefore, the light has a broad spectral range in the $x$ direction. Under these considerations, we assume the polaritonic dispersion to be,
\begin{equation}
    \mathcal{E}(\vec{k}) = \sqrt{\big((c/n_0)k_x\big)^2 + \big(v_gk_z\big)^2},
\end{equation}
where $v_g=\frac{\partial \omega_\text{LP}(k_z)}{\partial k_z}$ is the LP group velocity that represents the field propagation in $z$-direction, under strong resonant coupling to the exciton mode. For now, we neglect the dispersive curvature -- that is, it is assumed that the photonic field linewidth is sufficiently narrow as compared to the dispersive curvature. Similarly, we also neglect the $\vec{k}$-dependence of the LP nonlinearity $g$. In $x$ direction, in contrast, the light is broadband due to the transverse waveguide confinement and we can neglect the excitonic coupling, so that only the photonic linear dispersion (material speed of light $c/n_0$) matters. We consider the case $\gamma(\vec{r})=0$ -- no polaritonic losses are included in the simulations.

Eq.~\eqref{eq:GPE2D} is efficiently solved using the split-step method, where each differential time step the interaction and potential term are evaluated in real-space, while the dispersion $\mathcal{E}(\vec{k})$ is applied to the Fourier modes, obtained after a 2D fast Fourier transform (FFT).

\subsection{Results of 2D simulation}
\begin{figure*}[ht]
    \centering
    \begin{subfigure}[b!]{0.63\textwidth}
        \centering
        \includegraphics[width=\textwidth]{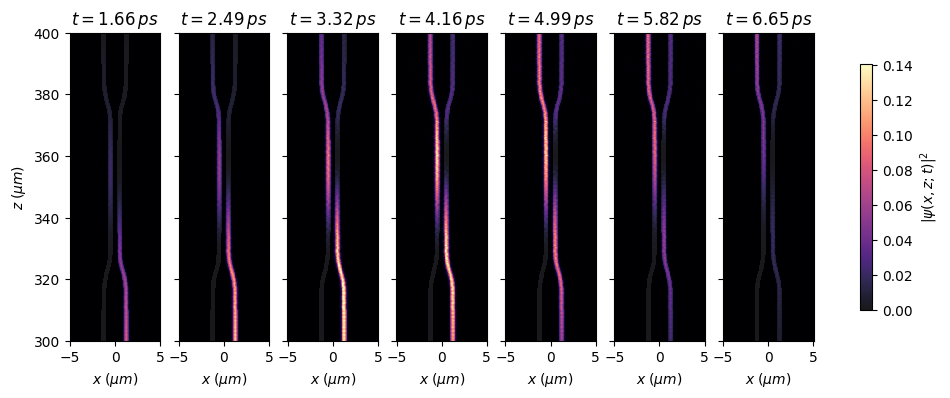}
        \caption{Time slices of 2D field simulations}
        \label{fig:subfig1}
    \end{subfigure}
    \hfill
    \begin{subfigure}[b!]{0.35\textwidth}
        \centering
        \includegraphics[width=\textwidth]{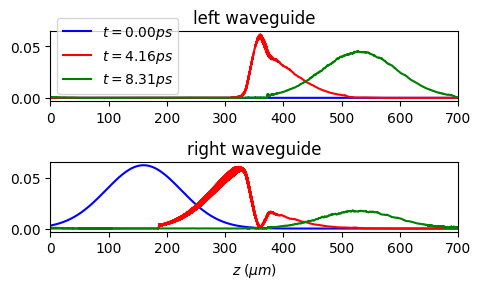}
        \caption{The extracted 1D waveguide fields}
        \label{fig:subfig2}
    \end{subfigure}
    \caption{Results of 2D LP field simulations of one nonlinear gate, with $50\,\mu m$ coupling. (a) Time slices of the 2D field density as the photon pulse propagates through. (b) The waveguide 1D field occupations at the initial (blue), middle (red) and final (green) simulated time steps, obtained by integrating over the $x$ direction.}
    \label{fig:GPE_2D}
\end{figure*}
We present the results of a 2D planar field simulation in Fig.~\ref{fig:GPE_2D}. The parameters used for the simulation are based on GaAs 2D planar heterostructures. In summary, we used $780$ nm input light, injected in waveguides of width $0.5\,\mu \text{m}$, with coupling length of $50\,\mu \text{m}$, at $0.5\,\mu \text{m}$ separation. The in-plane refractive index of the 2D sample is set to $n_0=3.28$, valid for a 2D stack of GaAs (quantum well) and AlGaAs layers (cladding) at cryogenic temperatures in the near-infrared regime, and we set $\Delta n=0.1$, realized by the etching in the cladding. The pulse duration is $\sigma_t=1\,\text{ps}$ and it contains an average of $\mathcal{N}=10$ photons, injected in the lowest transverse eigenmode of the waveguide potential ($\text{TE}_0$-mode). The nonlinearity was set to $g=600\, \mu eV\cdot \mu m^2$, in accordance with Ref. \cite{liran2024electrically}. Finally, the LP group velocity was $v_g = 0.5\cdot c/n_0$, so a factor $2$ lower than the photonic velocity, due to the strong excitonic coupling of the light. This corresponds to an exciton fraction of $u_k^2=0.5$. 

In Fig.~\ref{fig:GPE_2D}(a), time slices of the simulation are shown, while Fig.~\ref{fig:GPE_2D}(b) shows the calculated 1D waveguide field, obtained by integrating over the transverse direction $x<0$ (left waveguide) or $x>0$ (right waveguide). By comparison with the same simulation for a linear coupler, that is, with $g=0\,\mu \text{eV} \cdot \mu \text{m}^2$, we can identify the parameters that define this gate -- a linear coupling ratio of $0.72$, meaning $\theta=J\Delta t=0.68\pi$, and a gate accumulated nonlinearity phase shift of $\theta_0^{(\text{nl})}=0.14\pi$ on the left side and $\theta_1^{(\text{nl})} = 0.11\pi$ on the right. This was obtained for a gate extent of $400\,\mu m$ and a time $t_\text{end} = 8.31\,\text{ps}$. For realizing the coupling, a gate length of $100\,\mu \text{m}$ is sufficient, in line with the parameter estimate at the end of Sec.\ref{subsec:cat} of the main text. 

The gate nonlinearity rate $U$ can be evaluated from the injected normalized pulse field intensity,
\begin{equation}
\label{eq:pulse-shape}
    n(\vec{r}) = \frac{1}{\sqrt{2\pi} \sigma_z}\exp\bigg\{-\dfrac{(z - z_0)^2}{4\sigma_z^2}\bigg\} \times \mathcal{F}_\text{tr}(x),
\end{equation}
with $\sigma_z = v_g \sigma_t$ the spatial pulse extent and $\mathcal{F}_\text{tr}(x)$ the transverse confined eigenmode of the waveguide. From this we find,
\begin{eqnarray}
\label{eq:U_2D}
    U = \int_\vec{r} d^2r\,n(\vec{r}) \big[g\, n(\vec{r})\big] =: \frac{g}{S_\text{pulse}}
\end{eqnarray}
Using the parameters of simulation, we find a gate nonlinearity of $U=0.0065\,\text{ps}^{-1}$ and effective pulse surface $S_\text{pulse}=141\,\mu\text{m}^2$, so that $U\Delta t=0.013$, for a $100\, \mu \text{m}$ gate. Note that this can be further improved with a slower group velocity $v_g$; this will both extend the exposure time to the nonlinearity and contract the incoming pulse stronger in the direction of propagation. Therefore, a factor $4$ in reduction of $v_g$, will yield a factor $16$ of improvement for $U\Delta t$. This would bring us into the regime considered for, e.g., the case of cat-state optimization in Sec.~\ref{subsec:cat}, where values up to $U\Delta t$ were considered for $>100 \mu m$ gate structures. This could be achieved in experiment by exciting at high exciton fraction \cite{walker2013exciton} and using techniques of slow light \cite{torrijos2021slow}.

\section{Deriving the gate parameters under pulsed laser excitation}
\label{app:MF1D}
For the purpose of estimating more carefully the parameters that define the gates, we reduce the full 2D waveguide field simulation from Appendix \ref{app:MF2D} to a set of two coupled 1D equations, with effective coupling $J_{i,i\pm1}$ between neighboring waveguides,
\begin{widetext}
\begin{equation}
\label{eq:GPE}
    i\big(\partial_t + v_g\partial_z \big) \psi_i(z, t) = g_\text{1D} \big|\psi_i(z,t)\big|^2 \psi_i(z,t) - J_{i-1,i}(z,t)\,\psi_{i-1}(z,t) - J_{i,i+1}^\ast(z, t)\,\psi_{i+1}(z,t).
\end{equation}
\end{widetext}
Here, $\psi_i(z,t) = \langle \psi_\text{LP,i}(z,t)\rangle$ is the LP field in waveguide $i$ at time $t$ and position $z$ in the waveguide and $v_g$ the LP group velocity. The couplings $J_{i,i\pm1}(z,t)$ describe the tunneling rates between two adjacent waveguides, as can be estimated, to first order, from the waveguide $\text{TE}_{0}$-mode profile overlap \cite{haus2002coupled}. However, more accurate values can be obtained with a simulation of the full 2D gate architecture in the linear regime, as was presented in Appendix \ref{app:MF2D}.

The parameter $g_{1D}$ is the 1D polariton interaction constant, which is found from $g$ by integrating out the transversal dimension of waveguide confinement,
\begin{equation}
\label{eq:g1D}
    g_{1D}(z) = g \int dx\, n(x,z)^2,
\end{equation}
with $n(x,z)$ the normalized pulse shape from Eq.~\eqref{eq:pulse-shape}. We set $g_{1D}(z)=g_\text{1D}$ in the simulation, thus again neglecting the dispersive curvature for the pulse propagation.

\begin{figure}[!ht]
    \centering
    \begin{subfigure}[b]{\linewidth}
        \centering
        \includegraphics[width=\linewidth]{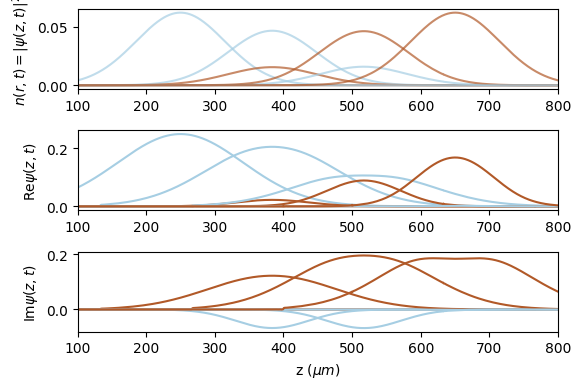}
        \caption{Snapshots of the pulse propagating from left to right, at $2.93\,\text{ps}$ apart}
        \label{fig:sub1}
    \end{subfigure}
    \hfill
    \begin{subfigure}[b]{\linewidth}
        \centering
        \includegraphics[width=\linewidth]{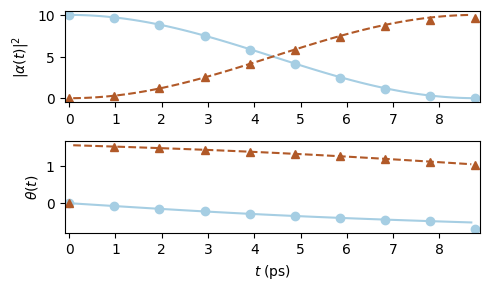}
        \caption{Comparing the 1D field dynamics (points) with the a circuit gate model (lines)}
        \label{fig:sub2}
    \end{subfigure}
    \caption{The propagation of a pulse through a $\pi/2$ swap gate with a strong photonic nonlinearity. The pulse is injected in first waveguide (blue) and transfers to the second (brown) (a) different snapshots, $\Delta t=2.93\,\text{ps}$ apart from each other, of the pulse propagating through the coupled waveguides from left to right; the photon density (up), and the real (middle) and imaginary part (bottom) of $\psi_{0,1}(z,t)$. (b) The mapping to an effective circuit model, using the circuit coupling $J$ and nonlinearity $U$. The values from the circuit simulation (full and dashed lines) are compared with the values extracted from the field simulation (dots and triangles), same color codes as (a).}
    \label{fig:circuit_mapping}
\end{figure}

A pulse of Gaussian shape, containing an average of $\mathcal{N}=10$ photons, is injected in the first waveguide ($i=0$) at position $z_0$ and width $\sigma_z = v_g \sigma_t$, with $\sigma_t$ the pulse duration,
\begin{equation}
\label{eq:pulse-in}
    \psi_0(z,t=0) = \sqrt{\frac{\mathcal{N}}{\sqrt{2\pi} \sigma}}\times e^{i k_z z} \times \exp\bigg\{-\dfrac{(z - z_0)^2}{4\sigma_z^2}\bigg\}.
\end{equation}
 The wavevector $k_z = n_0k_0$, with $n_0$ the effective 1D refractive index and $k_0=2\pi/\lambda$, with $\lambda$ the wavelength -- $\lambda=780\,\text{nm}$ for exciton resonance in GaAs.

Starting from the pulse 1D field dynamics, the goal is to establish a photonic circuit model of the form,
 \begin{eqnarray}
\label{eq:eff_ODE}
    i\partial_t \alpha_{0} &=& -J \alpha_{1} + U\big|\alpha_{0}\big|^2 \alpha_{0}, \\
    i\partial_t \alpha_{1} &=& -J \alpha_{0} + U\big|\alpha_{1}\big|^2 \alpha_{1}.
\end{eqnarray}
Here, $\alpha_i$ are bosonic field modes that describe the photonic component of the pulse. Furthermore, $J$ is the linear coupling between the modes and $U$ is the photonic nonlinearity, both expressed as temporal rates (inverse time units). 

The normalized 1D field pulse profile at incidence,
\begin{equation}
\label{eq:w}
    w(z, t=0) = 
    %\frac{\psi_0(z,t=0)}{\big(\int_{-\infty}^{\infty}|\psi_0(z,t=0)\big|^2 dz\big)^{\frac{1}{2}}} = 
    \frac{1}{\sqrt{\mathcal{N}}}\psi_0(z,t=0),
\end{equation}
serves as a reference for obtaining the bosonic circuit modes by evaluating the field overlap at later times $t$, in waveguide $l$,
\begin{equation}
\label{eq:a_pulse}
    \alpha_l(t) = \int_{-\infty}^{\infty} dz\; w(z, t=0) \,\psi_i(z,t).
\end{equation}
The circuit nonlinearity is derived similarly as for the 2D case \eqref{eq:U_2D}, 
\begin{equation}
\label{eq:U_pulse}
    U = g_{1D} \int_{-\infty}^{\infty} dz\;\big|w(z,t=0)\big|^4.
\end{equation}
As explained above for the 2D case, the circuit nonlinearity $U$ can be changed by changing the pulse duration $\sigma_t=\sigma_z/v_g$ (see Eq.~\eqref{eq:pulse-in}); shorter pulses give higher $U$. 

In Fig.~\ref{fig:circuit_mapping}, we show the simulation of two coupled waveguides (Fig.~\ref{fig:circuit_mapping}(a)) and the corresponding mapping to circuit gate model \eqref{eq:eff_ODE} ((Fig.~\ref{fig:circuit_mapping}(b)), in a regime of relatively strong nonlinearity, $g_{1D}=1200\,\mu\text{eV}\cdot\mu\text{m}$, similar to what was reported in \cite{liran2024electrically} in case of $\sim 0.5\,\mu\text{m}$ transverse waveguide confinement -- see parameters used in Appendix \ref{app:MF2D}. The linear gate coupling was set to represent a swap gate, $J\Delta t=\pi/2$, with $\Delta t$ the time of integration, and the pulse has a duration set to $\sigma_t=1\,\text{ps}$, containing on average $\mathcal{N}=10$ photons. The LP group velocity was set to $v_g= 45.58\,\mu\text{m}\cdot\text{ps}^{-1}$, corresponding to a reduction factor of about $2$ as compared to bare GaAs-based samples due to the exciton coupling (LP exciton fraction $u_k^2\approx0.5$, see Eq.~\eqref{eq:ex_ph_mixing_LP_UP} and below); this is in line with values reported in Ref. \cite{walker2013exciton}. The waveguide dynamics is shown in Fig.~\ref{fig:circuit_mapping}(a), where it is seen that the nonlinearity causes, in the first place, intensity-dependent phase shifts. In Fig.~\ref{fig:circuit_mapping}(b), a comparison with the mapping to circuit gate dynamics from Eqs.~\eqref{eq:eff_ODE} is shown, using the bosonic field modes $\alpha_{i}(t)$ defined in Eq.~\eqref{eq:a_pulse}. We derive an effective gate nonlinearity of $U=0.013\, \text{ps}^{-1}$ or, in length units, $U/v_g=0.38\,\text{mm}^{-1}$. Consequently, for a gate design of, e.g., $500\,\mu\text{m}$, we would find a dimensionless value of $U\Delta t\approx0.2$. 

In \ref{fig:circuit_mapping}(b), the interaction-induced phase shift that develops over time in the 1D field simulation of the pulses (dots and triangles) is in good agreement with the effective circuit simulation. However, when considering this regime of very strong photonic nonlinearity, small deviations start showing up at long times. This is attributed to the strong nonlinearity $g_{1D}$ and a high number of $\mathcal{N}=10$ photons per pulse, which causes in-pulse phase fluctuations (see final time slice \ref{fig:circuit_mapping}(a)) and therefore slightly affects the accuracy of the overlap integral from Eq.~\ref{eq:a_pulse}.

% \section{Gradients of singular value decomposition for complex matrices}
% \label{app:svd}
% The gradients towards the singular value decomposition (SVD) can be evaluated

\section{Quantum Fisher information in coherent input state}
\label{app:QFIcoherent}

The quantum Fisher information (QFI) is defined as,
\begin{equation}
\label{eq:QFI}
    \text{QFI}(\theta) = 4 \cdot \lim_{\theta\rightarrow 0} \frac{1-\mathcal{F}\big[\rho(0), \rho(\theta)]}{\theta^2},
\end{equation}
with,
\begin{equation}
    \mathcal{F}\big[\rho_1, \rho_2] = \Big(\text{Tr} \sqrt{\sqrt{\rho_1} \rho_2 \sqrt{\rho_1}}\Big)^2,
\end{equation}
the quantum Uhlmann fidelity between the unperturbed state $\rho(0)$ and the perturbed $\rho(\theta)$, which has a small phase shift $\theta$. It provides the maximal classical Fisher information that can be obtained from the measurement statistics of an optimal Hermitian observable $B$ \cite{polino2020photonic},
\begin{equation}
\label{eq:sup_QFI}
    \text{QFI}(\theta) = \sup_B \text{FI}\big(\theta | B\big),
\end{equation}
with,
\begin{equation}
    \text{FI}\big(\theta | B\big) = \sum_b P_b(\theta|B) \times \big( \partial_\theta \log P_b(\theta|B) \big)^2,
\end{equation}
where $P_b(\theta|B)=\langle b|\rho(\theta)|b\rangle$ are the classical probabilities of the measurement outcome $b$ of the Hermitian operator $B$, i.e., the eigenmodes $B|b\rangle = b|b\rangle$.

For the use case of quantum phase sensing considered in Sec.~\ref{sec:metrology}, a small phase shift was given to one circuit input mode, as expressed in Eq.~\eqref{eq:init} from main text. The QFI is obtained from the overlap between the unperturbed and perturbed state, 
\begin{eqnarray}
    \nonumber\mathcal{F}\big[\rho^{(l)}(\theta), \rho^{(l)}(0)\big] &=& |\langle \alpha e^{i\theta}|\alpha\rangle|^2\\
    \nonumber&=& e^{-|\alpha|^2|e^{i\theta}-1|^2} \\&=& 1-|\alpha|^2\theta^2 + \mathcal{O}(\theta^4).
\end{eqnarray}
Hence, from Eq.~\eqref{eq:QFI}, we find the quantum Fisher information contained in a phase-shifted coherent state to be $\text{QFI}=4|\alpha|^2$, directly proportional to the coherent photon intensity $|\alpha|^2$.

\section{Numerical analysis of MPS optimization for cat-state preparation}
\label{app:num_cat_state}

\begin{figure*}[!t]
    \centering
    \includegraphics[width=1\linewidth]{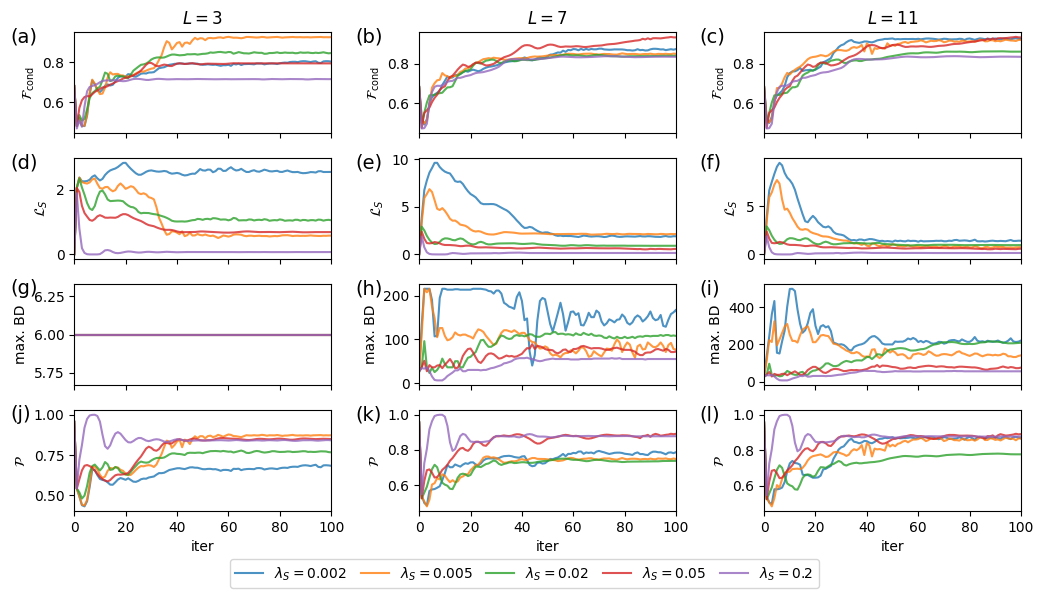}
    \caption{We show: (i) conditioned fidelity $\mathcal{F}_\text{cond}$ (a)-(c), (ii) entropy measure $\mathcal{L}_S$ from Eq. \eqref{eq:LS} (d)-(f), (iii) the max. bond dimension encountered in MPS (g)-(i) and (iv) the reduced-state purity $\mathcal{P}=\text{Tr}\rho^2$ for the cat-state attempt, during the number of iterations in optimization (x-axis) (j)-(l). We consider number of waveguides $L=3$ (left column), $L=7$ (middle) and $L=11$ (right). Of particular importance is the weight given to entanglement entropy penalty $\mathcal{L}_S$ during optimization, $\lambda_S$ (colors) -- $\lambda_S=0.005$ (blue) was used in Fig. \ref{fig:fig-cat} from the main text. We see that an intricate balance is needed between $\lambda_\rho$ (weighing the trace distance, set to unity) and $\lambda_S$, due to the strong non-convexity of the optimization problem.}
    \label{fig:convergence}
\end{figure*}

In this Appendix, we present an elaborate numerical analysis of the cat-state unitary-circuit optimization problem, especially concentrating on the role of the entanglement weight $\lambda_S$ from Eq. \eqref{eq:Ltot} in the main text. The goal of $\lambda_S$, chosen as $\lambda_S=0.005$ in main text, is twofold and can be seen as balancing search-space exploration and convergence stability. Although the final target cat-state must be pure single-mode (non-entangled), the full space of variational qPIC quantum states could, in principle, be explored to maximize state overlap with the final single-mode target. First, if $\lambda_S$ is too low, the couplings $J_{l,d}$ can grow in an uncontrolled manner, thus without restraint exploring the multimode photonic state-space during optimization. This comes with two immediate numerical considerations: (i) an uncontrolled growth of the MPS tensor bond dimensions, causing increases in circuit-evaluation times, and (ii) the drifting away from (local) optima due the strong non-convex and multivariate nature of the problem, coming with a large parameter space for exploration. Therefore, we want to attain control of the growth of $J_{l,d}$ during gradient optimization by adequately choosing $\lambda_S$; not too high, because in that case the state search-space will not be explored sufficiently to converge to a high-fidelity single-mode cat-state, and not too low, because in that case the numerics will become too heavy and there is no guarantee of accurate convergence.

In Fig. \ref{fig:convergence}, we illustrate this issue in full detail, for different waveguide number $L$, where we show the conditioned state fidelity $\mathcal{F}_\text{cond}$, the entropy measure for optimization $\mathcal{L}_S$, defined in Eq. \eqref{eq:LS} from the main text, the maximal quantum-state bond dimension reached and, finally, the purity $\mathcal{P}=\text{Tr}\rho^2$ of the cat-state attempt (which should be $\mathcal{P}=1$ for a pure state), for different values of the control parameter $\lambda_S$ (line colors) to control entanglement growth. Although $L=3$ and $L=11$ converged to nearly identical values $\mathcal{F}_\text{cond}=0.93$ for $\lambda_S=0.005$ in the main text (yellow lines, results from Fig. \ref{fig:fig-cat}), significant differences occur when $\lambda_S$ is altered. Seemingly, exactly the value $\lambda_S=0.005$ is a sweet spot for $L=3$, since other values, both lower and higher, result in lower fidelity $\mathcal{F}_\text{cond}$ (see Fig. \ref{fig:convergence}(a)). In that regard, the case with $L=11$ waveguides appears more robust toward $\lambda_S$, since the converged values $\mathcal{F}_\text{cond}$ lie closer to each other (Fig. \ref{fig:convergence}(c)). However, the converged fidelity outcomes after optimization are still not monotonous as a function of $\lambda_S$ -- see a sudden rise of $\lambda_S=0.05$ (red) in Fig. \ref{fig:convergence}(c), while $\lambda_S\in\{0.02,0.2\}$ (green, purple) are both lower.

In Fig. \ref{fig:convergence}(d)-(f), we illustrate the growth of entanglement during optimization, using the measure from Eq. \ref{eq:LS} in main text. During the first iterations, while the couplers $J_{l,d}$ increase from their initial values, a strong peak in entanglement entropy occurs, with an amplitude that depends directly on $\lambda_S$ (very clear in Fig. \ref{fig:convergence}(f)). This can be interpreted as a first phase of exploring the state-space, before converging to the final single-mode target cat-state, during which the redundant entanglement entropy is penalized. For the case $L=3$ (Fig. \ref{fig:convergence}(d)), it is observed that, on one side, a low value of $\lambda_S=0.002$ (blue line) does not significantly penalize the redundant entropy for forcing the convergence of fidelity, while, on the other, a high value $\lambda_S=0.2$ (purple line) immediately presses down any growth in entropy, without exploring the parameter space. Both cases result in poor converged values for $\mathcal{F}_\text{cond}$.

In Fig. \ref{fig:convergence}(g)-(i), the maximal bond-dimension (max. BD) in the tensor chain is shown under optimization. For low system sizes $L=3$ it always converges to $6$ for the SV truncation used ($s^2_\text{min}=10^{-12}$), as expected for the Fock-space dimension $d=6$ (max. $5$-photon occupation). For higher $L$, max. BD strongly grows for lower weights $\lambda_S$, since the waveguide bipartite entropy is less suppressed. As a reference, when determining the preferred version of $\lambda_S$, we tried to restrict to values of max. BD below $500$ during the optimization iterations -- this was just surpassed during the initial peak of entropy growth in Fig. \ref{fig:convergence}(i), for $L=11$ and $\lambda_S=0.002$ (blue line).

Finally, the converged attempted single-mode cat-state is not necessarily pure, but values around purity $\mathcal{P}\approx0.86$ are encountered, in the best cases -- see Fig. \ref{fig:convergence}(j)-(l). Thus, we conclude that the non-unit fidelity $\mathcal{F}_\text{cond}$ is, to some degree, a consequence of the non-purity of the converged qPIC cat-state attempt. Obviously, larger values of $\lambda_S$ lead to lower values of entanglement growth $\lambda_S$ and, hence, to high-purity (but low-fidelity) single-mode quantum states.

\begin{figure*}[!t]
    \centering
    \includegraphics[width=1\linewidth]{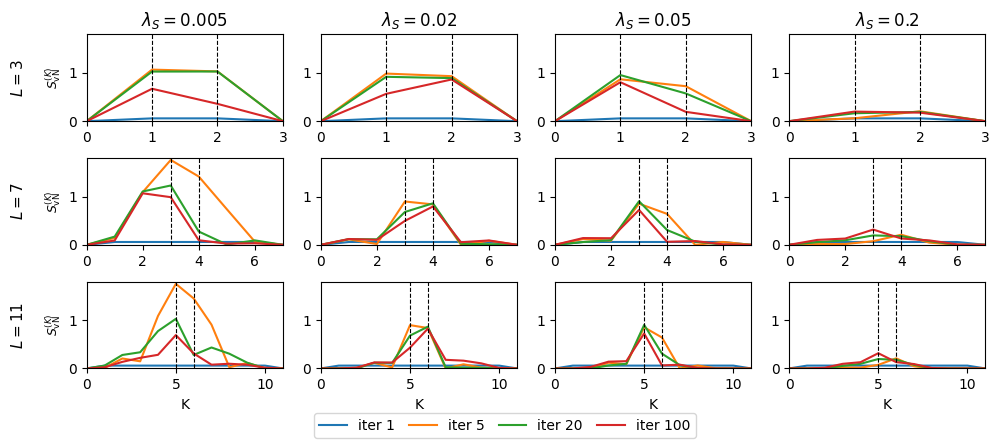}
    \caption{The inter-waveguide reduced-state bipartite entanglement entropy, defined in Eq. \eqref{eq:S_ent} from the main text, reached under cat-state optimization. Results are given for different waveguide number $L$ (rows) and entropy weights $\lambda_S$ (columns), with $k$ indicating bipartition in subsystems $l<k$ and $l\geq k$. The line color codes refer to the iteration in optimization -- iter 1 is the first circuit run, with small uniform $J_{l,d}$ and iter 100 the value equivalent to convergence. The marked vertical dashed lines are the bonds of importance for the reduced cat-state optimization. Larger values of $\lambda_S$ lead to larger suppression of buildup of entanglement entropy.}
    \label{fig:S_profile}
\end{figure*}

This final consideration is understood better when considering the full tensor chain bipartite entanglement-entropy profile, defined in Eq. \eqref{eq:S_ent} in the main text, as illustrated in Fig. \ref{fig:S_profile} for different values of $L$ (rows) and $\lambda_S$ (columns), different iterations under optimization (color codes). We have visualized the bonds of importance for the cat-state preparation objective (dashed vertical lines). As noted, the main part of the entanglement growth centers around the cat-state, for which we see, generally, a first stage of growth (orange-green lines) and, subsequently, a decline when the entanglement becomes more suppressed in the final convergence (red lines). Again we see that higher values of $\lambda_S$ (right columns) strongly suppress the buildup of bipartite entanglement.

\begin{figure*}[!t]
    \centering
    \includegraphics[width=1\linewidth]{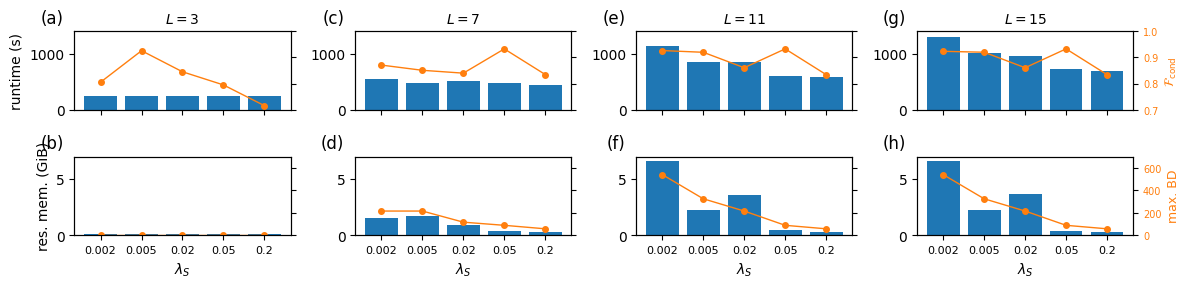}
    \caption{The scaling of runtimes and memory load.
    (a-c-e-g) Comparison between: cond. fidelity reached (orange dots) and computational runtime (blue bars)  and (b-d-f-h) reserved memory load (blue bars) and  max. BD reached (orange dots), for different mode number $L$ in 100-iteration cat-state optimization, under varying entropy weight $\lambda_S$. For illustration, we added the case $L=15$, not considered in main text. Interestingly, low values of $\lambda_S$ (left bars), requiring larger tensor bond dimensions, do not immediately cause an excessive increase in runtime as compared to higher values, by contrast, the increase memory load (blue bars, b-d-f-h) is the predominant concern.}
    \label{fig:runtimes}
\end{figure*}

In Fig. \ref{fig:runtimes} we illustrate the computational runtimes for a $100$-iteration optimization, under different values of $L$ (the columns left to right) and $\lambda_S$ (x-axis), and compare the values for the corresponding converged conditioned fidelity $\mathcal{F}_\text{cond}$ (orange dots, (a-c-e-g)), the max. BD reached (orange dots, (b-d-f-h)) and the reserved memory load, the dominant contributor for throwing an `out-of-memory' error in PyTorch (blue bars, (b-d-f-h)). We remind the reader that the optimizations were run on GPU (NVIDIA Tesla V100, 32 GiB RAM), using the PyTorch library (see code base \cite{coderepo}) which includes highly-parallel heavy-duty GPU efficiency. For $L=3$ (left), the runtimes are always equal, this is an obvious consequence of the tensor bond dimension being saturated at $\chi=6$ (panel (a-b)). For $L=7$ with $\lambda_S=\{0.002,0.005\}$, the bond dimension saturates at the maximal value, $d^3=216$, with the physical dimension $d=6$ (panel (c-d)). For $L=11$, we start marking a noticeable decay in runtimes with increased $\lambda_S$, due to the suppressed growth of bipartite entanglement and, hence, the required max. tensor bond dimension during optimization. Nevertheless, the computational overhead remains manageable for lower $\lambda_S$, despite the large variations in bond-dimensions (see in Fig.~\ref{fig:convergence}(i)). This is attributed to the massively parallel tensor operations executed on GPU via PyTorch, which result in efficient higher-rank tensor contractions and SVDs, with no significant overhead, provided the computation stays within GPU memory (32 GiB), avoiding costly host-device data transfers.

The reserved memory load under optimization immediately illustrates where problems start occurring. For $L=\{11,15\}$, with low entropy penalty, $\lambda_S=0.002$, we arrive at 7 GiB. Note the similarity in memory load between $L=11$ and $L=15$ (panel (f-h)). This is explained by the entropy penalty (Eq. \ref{eq:LS} in main text), which is not renormalized with $L$. From experience, we noticed that enlarging the physical dimension $d>6$ (higher than $5$ photon-truncation) soon started causing `out-of-memory' (OOM) error in PyTorch for large $L$.

For the note, all the simulations above were run with Adam learning rate $\text{lr}=0.1$ and started from the small (but non-zero) uniform value $J_{l,d}=J_\text{init}=0.01$.

To conclude this Appendix, we want to highlight the benefits of semi-exact MPS simulations for multimode photonic simulations, in the regime of few photon occupation per mode. One might reckon that a local Fock dimension $d=6$ for $L=11$ photonic modes leads to a full Hilbert space dimension of $\mathcal{D}=6^{11}\approx 3\cdot 10^8$ (about $5.41$ GiB in complex 128-bit representation), which would still be suitable to be handled with exact simulation dynamics using sparse matrix operations and dense vector state-representation. Although that may be true for a single run, when using highly optimized hardware, there are two important considerations: (i) multiple iterative runs for the optimization ($\sim100$ iterations) are needed to reach convergence and (ii) the gradients of \emph{any} gate operation must be tracked under optimization, with comes with the temporary storage of intermediate states. Especially the latter is important for memory considerations. The large part of the operations consist of gate contractions (linear, but gate tensors must be stored in memory for back-propagation of the tensor gradients) and SVDs. For efficiency, we even had to rewrite the PyTorch module to execute the truncated SVD scheme efficiently (only the truncated tensors need to be stored for gradient back-propagation) with support for complex-valued tensors (this requires additional terms in gradient backward pass \cite{wan2019automatic}). For comparison, we ran a quick test run using exact dynamics, for $L=9$ modes, $D=25$ layers, and arrived at $550$s simulation for a \emph{single} forward circuit run, on CPU, without gradient tracking, simulations for $L=11$ did not finish within the waiting time ($>1$h). By contrast, in Fig. \ref{fig:runtimes}, the highest runtime encountered was around $1200$s (right panel), for a total of $100$ iterations (i.e., circuit evaluation plus gradient backpropagation) and $L=15$ modes -- thus, a substantial gain in efficiency.

\section{Circuit robustness for odd cat state preparation}
\label{app:robustness}
\begin{figure*}[!ht]
    \centering
    \includegraphics[width=0.7\linewidth]{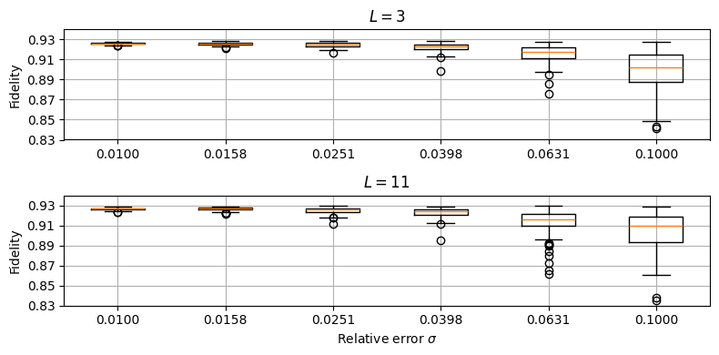}
    \caption{The robustness analysis for cat-state generation with deep circuits, $D=25$ and $U\cdot T_\text{circ}=6.25$, for $L=3$ (top) and $L=11$ (bottom) coupled waveguides, using boxplots.}
    \label{fig:robustness}
\end{figure*}

After establishing an optimal qPIC configuration for some optimization task, it is important to analyze the robustness and stability of the qPIC towards small fabrication errors and inaccuracies. In this appendix, we perform a circuit robustness analysis for odd cat-state preparation, as elaborated in Sec.~\ref{subsec:cat}, towards fabrication errors in the gate couplings. These can be caused by small imperfections in the width of the waveguides, the etch depth or the waveguide thickness, all of which will affect the accuracy of the target gate inter-mode coupling. Nevertheless, given the relatively low-index-contrast confinement in GaAs waveguides (limited etch depth, constrained by the QW), which grants greater error robustness, we assume waveguide width, and correspondingly effective waveguide separation, will dominate the fabrication noise \cite{haus2002coupled}.

Although the exact range of deviations in terms of those imperfections heavily depends on the maturity of the GaAs fabrication process \cite{chang2018heterogeneously,albiladi2022gaas,1088430}, we will perform a first phenomenological assessment of the impact of gate coupling errors, using the common assumption that these geometry deviations follow a Gaussian distribution \cite{melati_melloni_weng_daniel_2016,9962782}. Importantly, the fabrication noise is chosen multiplicative, mainly for two reasons. First, under evanescent mode coupling, based on the corresponding overlap integrals of the adjacent waveguide modes, the coupler value $J^\text{(opt)}_{l,d}\Delta t$ (see Eq. \ref{eq:V2}) depends exponentially on the waveguide edge separation (the photonic transverse mode-overlap). This leads to a log-normal distribution for $J^\text{(opt)}_{l,d}$, as a consequence of the Gaussian distribution of the geometry deviations. In the small noise limit, in first approximation, a multiplicative noise model can be used. Second, the coupling values $J^\text{(opt)}_{l,d}\Delta t$ can, by definition, not be negative-valued under evanescent coupling \cite{haus2002coupled, xing2017behavior,Somekh1973Channel}. Thus, when $J^\text{(opt)}_{l,d}$ is close to $0$ (large waveguide separation), small displacements in the waveguide edge positioning will not significantly change the coupling. The sensitivity is much stronger for stronger coupling, e.g., close to 50:50 coupling, which has $J^\text{(opt)}_{l,d}\Delta t=\tfrac{\pi}{4}$. 

Notably, we noticed in the literature the noise is often chosen additive instead (see, e.g., Refs. \cite{ullrick2026design,kosmella2026robustness}). However, the onset is often calibration of the robustness around a fixed coupling rate, such as 50:50, while we, in contrast, encounter a wide range of coupler ratios $J^\text{(opt)}_{l,d}\in[0,\tfrac{\pi}{2}]$, which requires appropriate rescaling of the noise caused separation mismatches.

In more concrete terms, for this study we applied multiplicative noise sampling around the optimal coupling rates $J^\text{(opt)}_{l,d}$,
\begin{equation}
\label{eq:robustness-noise}
    \tilde{J}^{(b)}_{l,d} \leftarrow J^\text{opt}_{l,d}(1 + \sigma \xi^{(b)}_{l,d}),
\end{equation}
with $\sigma$ the variance and $\xi_{l,d}^{(b)}\sim \mathcal{N}(0,1)$ normal-distributed random noise, the index $b$ labels the sample set. {\color{blue}  We scan the range of relative deviations $\sigma$ up to $10\%$, which we consider a reasonable number given the reported deviations in literature for similar GaAs waveguide platforms \cite{chang2018heterogeneously,albiladi2022gaas,1088430}, and more mature silicon-on-insulator waveguide platforms \cite{Xing2023}.}

In Fig.~\ref{fig:robustness}, we show the results for different levels of noise in the couplers (logarithmic scale), for deep circuits $D=25$ and a circuit nonlinearity $U\cdot T_\text{circ}=6.25$. We studied the robustness for $L=3$ and $L=11$ waveguides, which showed similar outcomes for fidelity after optimization in the text -- see Fig.~\ref{fig:fig-cat}(e)-(f). The box of each data point was generated with $100$ different random samples $b$, as defined in Eq.~\eqref{eq:robustness-noise}. Here again, there is no compelling evidence found for preferring larger circuits, with $L=11$, over smaller, with $L=3$, when considering their robustness. We see that odd cat-state generation is relatively stable towards fabrication errors in the gate couplings up to relative errors $\sigma\approx10\%$; this results in a loss of $2-3\%$ in average fidelity and an increase in spread (interquartile range) of about $8\%$, excluding outliers.

In summary, we performed an analysis of the circuit robustness towards fabrication errors in the couplers $J_{l,d}$ of qPIC optimized for odd cat-state generation. In the future, also detector noise, especially important for evaluating the photonic Fisher information, can be included in the in the robustness analysis, as well as imperfections related to in- and out-scattering of the qPIC. Furthermore, qPIC robustness can be incorporated as explicit requirement during the optimization, as was considered, for example, in Refs. \cite{jiang2024adeptzzeroshotautomatedcircuit,kosmella2024noise,gu2022} for neuromorphic computing. After outlining the relevant noise channels for a specific qPIC task, similar methods can be used in the tensor-network simulations.

\section{Gaussian quantum state sensing}
\label{app:gaussian_metrology}
\begin{figure}[!ht]
    \centering    
    \includegraphics[width=1\linewidth]{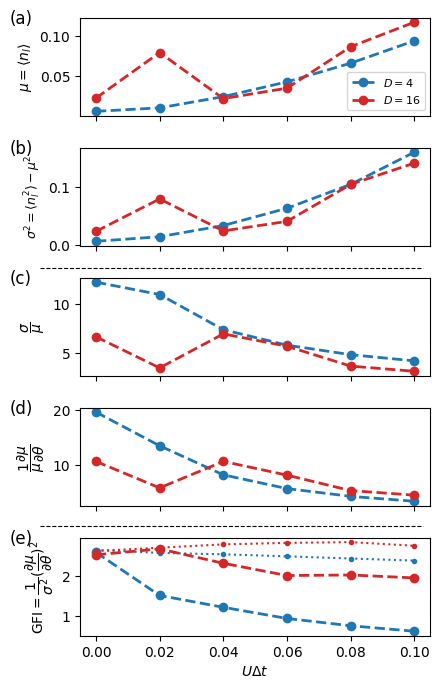}
    \caption{The Gaussian comparison for sensing, for $L=5$ waveguides, with $D=4$ (blue) and $D=16$ (red) circuit layers. (a) the mean photon number and (b) number the variance. (c) The variance $\sigma$ normalized with $\mu$ becomes lower for higher gate nonlinearity $U\Delta t$ and (d) the mean derivative towards $\theta$, normalized with $\mu$, both characterizing the output sensitivity. (e) The Gaussian Fisher information extracted for the experiment (dashed line), compared with the full Fisher information obtained after optimization in main text (dotted). While Gaussian FI contains a large part of the FI, the non-Gaussian statistics of the light are crucial for achieving a benefit over the classical linear case ($U\Delta t=0$).}
    \label{fig:gaussian_sensing}
\end{figure}
In the main text, Sec.~\ref{sec:metrology}, a qPIC optimization scheme was developed for optimal phase-shift sensing, based on detecting the photon number distribution of the one output mode. Here, we explain that, to some extent, the extraction of information from the circuit can be understood from Gaussian sampling methods. For this, we investigate the circuit sensitivity by means of the Gaussian Fisher Information (GFI),
\begin{eqnarray}
\label{eq:GFI}
    \text{GFI}:=\frac{1}{\sigma^2}\bigg(\frac{\partial \mu}{\partial \theta}\bigg)^2,
\end{eqnarray}
with, in the case of qPIC output signal sampling, the mean output of mode $l$, $\mu=\langle n_l\rangle$, and the variance, $\sigma^2=\langle n_l^2\rangle - \mu^2$. This coincides with standard formulation of Fisher information contained in Gaussian distributions, on the condition that $\frac{\partial \sigma^2}{\partial \theta}$ is negligible -- see, e.g., Ref. \cite{kay1993fundamentals}. 

In Fig.~\ref{fig:gaussian_sensing}, we present the results of the analysis. Fig.~\ref{fig:gaussian_sensing}(a)-(b) show the $\mu$ and $\sigma^2$, extracted from the number distribution, for circuits of $L=5$ waveguides and circuit time $T_\text{circ}=D\cdot \Delta t$, with $D=4$ (blue lines) and $D=16$ (red lines). 

In Fig.~\ref{fig:gaussian_sensing}(c)-(d), the standard deviation $\sigma$ and $\frac{\partial \mu}{\partial \theta}$ are shown, normalized with $\mu$. Here, it is observed that the standard deviation is generally lower for deeper circuits that have the same gate nonlinearity $U\Delta t$, indicating a small amount of amplitude squeezing at higher gate nonlinearity $U\Delta t$ and circuit depth $D$. In addition to that, the sensitivity of the mean $\frac{\partial \mu}{\partial \theta}$ becomes higher above $U\Delta t\approx 0.04$, but still lower than the linear case $U=0$. 

These two finding are apparent in Fig.~\ref{fig:gaussian_sensing}(e), where GFI is shown (dotted lines) and compared with the actual FI reached after qPIC optimization (Fig.~\ref{fig:sensing}(b), main text). Interestingly, for $U=0$, linear qPICs, all points coincide. That is, the FI extracted from the readout of one mode's photon number distribution $P^{(l)}_n$ is equivalent to the Gaussian FI, independent of circuit time $T_\text{circ}=D\cdot \Delta t$. This changes for higher $U\Delta t$; shallow circuits show an immediate decline in FI, due to initial effects of blockade, thus affecting the GFI that can be extracted from the output. When analyzing the full number distribution $P^{(l)}_n$, as in the main text, a much slower decline is observed (blue dotted line), which we attribute to the non-Gaussian nature of the photon statistics. In the case of a deep circuit with $D=16$, this changes somewhat. At higher nonlinearity $U\Delta t$, the GFI declines as well, but much slower than the case $D=4$, due to the stronger potential to exploit the multimode interferences. As we found in main text, the number-distribution FI contained in $P_n$ slowly increases for small $U\Delta t$ and finds a maximum at $U\Delta t=0.08$, before it declines again. Therefore, adding non-Gaussian elements in the statistics is, in the analysis of the Fisher information associated with number-resolved measurements, essential to observe the actual advantage of the qPIC compared to a PIC (linear qPIC).

It must be noted that, in the main text, the optimization was carried out for the FI defined on the full photon number distribution $P^{(l)}_n$ as explicit FOM -- see Eq.~\eqref{eq:FI}. When, on the other hand, the GFI itself is set as the FOM for optimization, this observation could change. Optimization schemes based on the Gaussian form for the FI, defined in Eq.~\eqref{eq:GFI} can be investigated in the future -- also motivated by the fact that this quantity would be more directly accessible in real-world experiments.

Furthermore, we see in Fig.~\ref{fig:gaussian_sensing}(a) that the signal intensity $\mu$ for optimal FI extraction becomes very low in the limit of small $U\Delta t$; $\mu=0.007$ and $\mu=0.023$, for $D=4$ and $D=16$, respectively, at $U=0$. By solely optimizing the FI \emph{an sich}, we see that the signal intensity already grows up to $\mu\approx 0.1$ for $U\Delta t=0.1$. Beyond that, also the signal intensity can be added explicitly in the FOM to avoid the decline in signal strength in the circuit output, which might affect the signal-to-noise ratio in experiment. This is similar to what was done for single-photon generation in Sec.~\ref{subsec:single-photon} and can be realized, e.g., by adding a term $\mathcal{L}_\text{signal} \propto -\langle n_l\rangle$ in the FOM for sensing defined in Eq.~\eqref{eq:Lps}.

\section{Homodyne readout limit for phase sensing sensing}
\label{app:homodyne-sensing}

The standard method for obtaining phase information from a photonic signal (classical or quantum) is homodyne detection, using one single linear beam splitter to interfere the signal with a reference beam (a local oscillator).

Generally, the output of one end of the beam splitter is given by,
\begin{eqnarray}
\label{eq:homodyne}
    \beta &=& c \,\alpha_s + e^{i\phi}\sqrt{1-c^2} \,\alpha_\text{LO}\\
    &=& c \,|\alpha_s|e^{i\theta} + e^{i\phi}\sqrt{1-c^2} \,|\alpha_\text{LO}|
\end{eqnarray}
with $\alpha_s$ the phase-shifted signal to be analyzed, $\alpha_{LO}$ the local oscillator, serving as reference, and $c^2$ is the mixing ratio and $\phi$ the relative phase. In the second step, it is assumed that $\alpha_s$ and $\alpha_\text{LO}$ are in phase, up to a small phase shift $\theta$ (global phase shifts do not matter).

The absence of photonic nonlinearities implies that the output signal of the beam splitter preserves the coherent Poisson photon-number statistics. For a Poisson distribution with mean $\mu$, the Fisher information is,
\begin{eqnarray}
   \text{PFI}(\theta)= \frac{1}{\mu}\Big(\frac{\partial \mu}{\partial \theta}\Big)^2,
\end{eqnarray}
equivalent to Eq.~\eqref{eq:GFI}, if $\sigma^2\equiv \mu$, as holds for Poissonian statistics. 

In our case, the mean $\mu$ is given by the signal intensity, so $\mu\equiv |\beta|^2$. Straightforward calculation leads to,
\begin{widetext}
    \begin{eqnarray}
        \text{PFI}(\theta)\Big|_{\theta=0} = \frac{4\sin^2(\phi)\,c^2\left(1-c^2\right)\,|\alpha_s|^2|\alpha_\text{LO}|^2}{c^2|\alpha_s|^2 +(1-c^2)|\alpha_\text{LO}|^2 + 2\cos(\phi)\, c\sqrt{1-c^2}\,|\alpha_s||\alpha_\text{LO}|}
    \end{eqnarray}
\end{widetext}
For the case of symmetric beam splitters, the couplers considered in the main text, we have that $\phi=\frac{\pi}{2}$. Furthermore, maximal mixing, $c=1/\sqrt{2}$ (balanced homodyne measurement), will give optimal results. In this case, we find,
\begin{equation}
\label{eq:symmFI}
    \text{PFI}(\theta)= 2\cdot\frac{|\alpha_s|^2|\alpha_{LO}|^2}{|\alpha_s|^2 + |\alpha_\text{LO}|^2}.
\end{equation}
For the low photonic occupation regime considered in Sec.~\ref{sec:metrology} of main text, $\alpha_s=\alpha_\text{LO}=1$, this leads to $\text{PFI}(\theta)=1$, significantly below the value found after circuit optimization -- see Fig.~\ref{fig:sensing} or Fig.~\ref{fig:gaussian_sensing}.

When we consider $|\alpha_\text{LO}|^2\gg|\alpha_s|^2$ in Eq.~\eqref{eq:symmFI}, the FI contained in one beam-splitter output converges to $\text{PFI}(\theta)=2|\alpha_s|^2$, still well below the results found after circuit optimization for reading out one output. Nevertheless, when \emph{both} outputs of the beam splitter are analyzed, with a strong local oscillator $\alpha_\text{LO}$, a perfect phase measurement is performed. In that case, an additional factor $2$ is gained and the Heisenberg limit, derived in Appendix \ref{app:QFIcoherent}, is recovered; $\text{PFI}(\theta)\equiv \text{QFI}(\theta)=4|\alpha_s|^2$.